\documentclass[prd,aps,showpacs,nofootinbib,eqsecnum,12pt]{revtex4}
\usepackage[dvips]{graphicx}

\usepackage{amssymb,amsmath}

\begin{document}


\title{Study of the preheating phase of chaotic inflation}

\author{Paul R. Anderson$^{\rm a}$}
\altaffiliation{\tt anderson@wfu.edu}

\author{Carmen Molina-Par\'{\i}s$^{\rm b}$}
\altaffiliation{\tt carmen@maths.leeds.ac.uk}

\author{David Evanich$^{\rm a}$}

\author{Gregory B. Cook$^{\rm a}$}
\altaffiliation{\tt cookgb@wfu.edu}

\affiliation{${}^{\rm a}$
Department of Physics,
Wake Forest University, Winston-Salem,
NC 27109, USA}

\affiliation{${}^{\rm b}$
Department of Applied Mathematics, School of Mathematics,
University of Leeds,
Leeds LS2 9JT, UK}


\begin{abstract}

Particle production and its effects on the inflaton field are
investigated during the preheating phase of chaotic inflation using a
model consisting of a massive scalar inflaton field coupled to $N$
massless quantum scalar fields.  The effects of spacetime curvature
and interactions between the quantum fields are ignored.  A large $N$
expansion is used to obtain a coupled set of equations including a
backreaction equation for the classical inflaton field.  Previous
studies of preheating using these equations have been done.  Here the
first numerical solutions to the full set of equations are obtained
for various values of the coupling constant and the initial amplitude
of the inflaton field.  States are chosen so that initially the
backreaction effects on the inflaton field are small and the mode
equations for the quantum fields take the form of Mathieu equations.
Potential problems relating to the parametric amplification of certain
modes of the quantum fields are identified and resolved.  A detailed
study of the damping of the inflaton field is undertaken.  Some
predictions of previous studies are verified and some new results are
obtained.

\end{abstract}

\maketitle

\section{Introduction}
\label{intro}

In all inflationary models there must be a period after inflation in
which a substantial amount of particle production occurs in order to
repopulate the universe with matter and radiation.  At the end of
this process it is crucial that the universe reheats to a
temperature that is not too large and not too
small~\cite{brandenberger}. If the temperature is too large, there are
problems with the creation of monopoles and domain walls that have not
been observed, and if the temperature is too low there can be problems
with baryogenesis~\cite{baryogenesis}.  Thus, to assess the viability
of a particular model of inflation, it is necessary to obtain an
accurate picture of the amount of particle production that occurs and
the rate that it occurs at.  This phase of particle production after
inflation is usually called reheating~\cite{reheating,kofman}.

Both the mechanism for inflation and the way in which reheating occurs
are different for different inflationary models.  In both new
inflation~\cite{new} and chaotic inflation~\cite{linde}, inflation is
due to a classical scalar field called the inflaton field.  When
inflation is over, this field oscillates in time as it effectively
``rolls back and forth'' in its
potential~\cite{linde,kolb-turner,brandenberger1}.  The coupling of
this classical field to various quantum fields and in some cases to
its own quantum fluctuations, results in the production of particles
which are thermalized by their mutual interactions.  The backreaction
of the quantum fields on the classical inflaton field causes its
oscillations to damp~\cite{brandenberger}.

The original particle production calculations for new inflation and
chaotic inflation~\cite{new-reheating} were done using perturbation
theory.  The damping of the inflaton field was taken into account
through the use of a dissipative term that was added to its wave
equation.  Later, the mode equations for the quantum fields coupled to
the inflaton field were solved and it was found that the time
dependent part of the modes can, in some cases, undergo parametric
amplification due to the periodic behavior of the inflaton
field~\cite{kls1,kls2,chaotic-reheating}.  Since the rate of particle
production is extremely rapid in this case and the particles do not
have time to thermalize while it is occurring, this initial phase of
the particle production process was dubbed ``preheating''~\cite{kls1}.

The process of preheating is interesting because particle production
occurs very rapidly with the result that quantum effects are large and
the backreaction on the inflaton field is important.  This pushes the
semiclassical approximation, which is what is typically used in these
calculations, to the limit.  In fact, to take such large effects into
account it is necessary to go beyond the ordinary loop expansion.  The
Hartree approximation and the large $N$ expansion are two ways in
which such effects have been treated~\cite{boy1,bc}.

The original calculations of preheating~\cite{kls1,kls2} involved
solving the mode equations for the quantum fields in the presence of a
background inflaton field which obeyed a classical scalar field wave
equation. Later estimates of the effects of the quantized fields on
the inflaton field were made using a variety of
techniques~\cite{KLS,boy0,boy1-no-O(N),boy-desitter,boy2,finelli}.  In
Refs.~\cite{tkachev,prokopec-roos} the first fully non-linear
calculations of classical inflaton decay were carried out by means of
statistical field theory lattice numerical simulations.  The numerical
simulations in~\cite{tkachev} and later ones
in~\cite{felder-kofman,kt3,DFKL,pfkp,dfkpp,fk1} were implemented by
means of the lattice code Latticeasy~\cite{latticeasy}. These
simulations are important for the study of the development of
equilibrium after preheating, when scattering effects cannot be
neglected~\cite{felder-kofman}. Recently, the same type of calculation
has been used to investigate the possibility that non-equilibrium
dynamics, such as that of parametric resonance during preheating, can
produce non-Gaussian density perturbations~\cite{chambers}.

The full set of coupled equations relating the inflaton field to the
quantum fields were solved numerically for both the Hartree
approximation with a single massive scalar field with a quartic
self-coupling and the $O(N)$ model using a large $N$ expansion
truncated at leading order.  The $O(N)$ model consists of $N$
identical massive scalar fields each with a quartic self-coupling and a quadratic coupling to every other field.
As a result there are $N$
inflaton fields in this model.  The first set of calculations were
done in a Minkowski spacetime background~\cite{boy1,boy3} and then
followed by radiation and matter dominated Friedmann-Robertson-Walker
backgrounds~\cite{boy4}.  As the fields have an energy-momentum tensor
associated with them, the expansion of the universe is also affected.
This was taken into account in Ref.~\cite{ramsey-hu}.  A numerical
calculation in the context of the SU(2) Higgs model was done in
Ref.~\cite{baacke}.  A calculation within the context of the $O(N)$
model that takes scattering of the produced particles into account was
done in Ref.~\cite{berges-serreau}.  Numerical calculations in models with
two scalar fields, with backreaction effects taken into account using the Hartree
approximation, were done in Ref.~\cite{Jin-Tsu}.

There are some potential problems with using the $O(N)$ model as a
model for chaotic inflation.  One is that it results in $N$ inflaton
fields rather than the single inflaton field that is usually
postulated.  Another is that the fields have a quartic self-coupling
as well as couplings to each other.  There is evidence from the WMAP
III data~\cite{inflaton-potential} that if chaotic inflation occurred,
the inflaton field is a massive scalar field with no quartic
self-coupling.  Finally, in the $O(N)$ model or in a model in which
the primary quantum fluctuations come from the inflaton field, the
quantum fields are massive with the same mass as the inflaton
field.  However, in a realistic situation there would also be massless
or effectively massless fields coupled to the inflaton field.

A possibly more realistic model for chaotic inflation consists of a
single classical inflaton field coupled to various quantum fields,
including its own quantum fluctuations.  This is the type of model
considered by Kofman, Linde, and Starobinsky~\cite{kls1,kls2,KLS}.
They used solutions of the mode equations of the quantum fields to
estimate both the amount of particle production that would occur in
various cases and the amount of damping of the inflaton field that
occurs due to the backreaction of the quantum fields on it.

In this paper we consider a model similar to one used by Kofman,
Linde, and Starobinsky, henceforth referred to as KLS, in their
analysis of preheating in chaotic inflation~\cite{KLS}.  The inflaton
field is a classical massive scalar field with minimal coupling to the
scalar curvature.  It has no quartic or higher order self-coupling.
But it is coupled to $N$ identical massless quantized scalar fields
with arbitrary curvature coupling $\xi$.  As discussed in Section II,
in the large $N$ limit, the system effectively reduces to one
consisting of a classical inflaton field coupled to one quantized
massless scalar field with arbitrary curvature coupling.  The effects
of quantum fluctuations of the inflaton field come in at next to
leading order so we do not take them into account.

We investigate some of the details of the preheating process by
numerically integrating both the mode equations for the quantized
field and the backreaction equation for the inflaton field.  These are
the first solutions to the full set of coupled equations relating the
inflaton field to the quantum field that have been obtained for this
model.  To simplify the calculations we ignore interactions between
the created particles. Calculations that have taken such interactions
into account~\cite{tkachev,prokopec-roos,berges-serreau,pfkp,dfkpp}
indicate that they can be ignored during the first stages of
preheating but eventually become important.  Another simplification is
that, as a first step, we work in a Minkowski spacetime background.
It has previously been shown that the expansion of the universe can
significantly affect the evolution of the inflaton field during
preheating if this field is of Planck scale or larger at the onset of
inflation~\cite{KLS}.  This in turn can substantially affect the
details of the particle creation process.  Thus, some of our results
will not be relevant for most models of chaotic inflation.  However,
for the rapid damping phase which occurs in many models, the amplitude
of the inflaton field changes rapidly on time scales which are small
compared to the Hubble time.  Thus, it should be possible to ignore
the expansion of the universe during this phase~\cite{KLS}.

If, for the model we are considering, one neglects the backreaction
effects of the quantized fields on the inflaton field, then the
inflaton field undergoes simple harmonic motion with a frequency equal
to its mass.  As shown in Section III, this means that the time
dependent part of the modes of the quantized field obey a Mathieu
equation.  Thus, parametric amplification will occur for modes in
certain energy bands.  There is an infinite number of these bands and
some occur for arbitrarily large values of the energy.  This could in
principle result in divergences of quantities such as $\langle \psi^2
\rangle$, with $\psi$ the quantum field.  For example, if the
contribution to the mode integral in $\langle \psi^2 \rangle$ [see
  Eq.~\eqref{bare-psi-squared}] for each band of modes undergoing
parametric amplification was the same at a given time, then $\langle
\psi^2 \rangle$ would diverge.  Another issue that must be addressed
is that even if $\langle \psi^2 \rangle$ is finite, one must be
certain that all significant contributions to it are accounted for
when making numerical computations; in other words, one must be
certain that no important bands are missed.  These issues are
addressed in Section III.

When the backreaction of the quantized fields is included in the wave
equation for the inflaton field, then, as expected, the inflaton
field's amplitude damps as the amplitudes of some of the modes of the
quantized field grow.  KLS predicted that if all the instability bands
are narrow then the damping is relatively slow while if one or more of
the instability bands are wide then there is a period of rapid
damping.  We find this to be correct, and for a Minkowski spacetime
background, we find a more precise criterion which determines whether
a phase of rapid damping will occur.  In cases where there is rapid
damping, we find a second criterion that must be satisfied before it
takes place.  In a Minkowski spacetime background, this latter
criterion explains the differences that are observed in the time that
it takes for the period of rapid damping to begin.

In their study of the rapid damping phase, KLS gave a criterion for
when the damping should cease and used it to predict how much damping
should occur.  We find something similar except that the amount of
damping that is observed to occur is larger than they predict.  We
also find that the rapid damping actually occurs in two phases separated
by a short time.  An explanation for this is provided in
Section~\ref{results-details}.

If most of the damping occurs gradually then it is observed that the
frequency of the oscillations of the inflaton field changes slowly as
its amplitude is damped.  There is also a significant transfer of
energy away from the inflaton field as would be expected.  However, if
most of the damping occurs rapidly then during the rapid damping phase
the frequency of the oscillations of the inflaton field increases
significantly.  As a result there is less energy permanently
transferred away from the inflaton field than might be otherwise
expected.  This result was seen previously in the classical lattice
simulation of Prokopec and Roos~\cite{prokopec-roos}.

No significant further damping of the amplitude of the inflaton field
was observed to occur after the rapid damping phase in all cases in
which such a phase occurs.  Instead both the amplitude and frequency
of the inflaton field were observed to undergo periodic modulations
and a significant amount of energy was continually transferred away
from and then back to the inflaton field.  However the actual
evolution of the inflaton field after the rapid damping phase is
likely to be very different because interactions
that we are neglecting are expected to be important during this
period~\cite{felder-kofman,tkachev,prokopec-roos,berges-serreau,pfkp,dfkpp}.

In Section II the details of our model are given and the coupled
equations governing the inflaton field and the quantum fields are
derived.  In Section~\ref{parametric} a detailed study is made of
parametric amplification and the bands that undergo parametric
amplification in order to address the issues discussed above relating
to the finiteness of certain quantities and the potential accuracy of
numerical computations of these quantities.  In Section IV some of our
numerical solutions to the coupled equations governing the modes of
the quantum field and the behavior of the inflaton field are presented
and discussed.  Our results are summarized in Section V.  In the
Appendix we provide the details of the renormalization and covariant
conservation of the energy-momentum tensor for the system under
consideration.  Throughout this paper we use units such that $\hbar =
c = G = 1$.  The metric signature is $(- +++)$.


\section{The Model}
\label{twofield}

We consider a single inflaton field $\Phi$ and $N$ identical scalar
fields $\Psi_j$, that represent the quantum matter fields present
during the inflationary phase. We assume that the inflaton field is
massive (with mass $m$) and minimally coupled to the scalar curvature,
$R$, and that the quantum scalar fields are massless with arbitrary
coupling $\xi$ to the scalar curvature. We also assume that the
interaction between the classical inflaton and the quantum matter
fields is given by ${g^2}\Phi^2 \Psi_j^2/2$.  We study the dynamics of
$\Phi$ and $\Psi_j$ in a Minkowski spacetime background and denote the
metric by $\eta_{\mu\nu}$.

The action for the system is given by
\begin{eqnarray}
S[\Phi, \Psi_j]&=&
-\frac{1}{2}  \int {\rm d}^4 x
\left(
\eta^{\mu \nu} \nabla_\mu  \Phi \nabla_\nu \Phi
+ m^2 \Phi^2
\right)  \nonumber \\  &  &  \nonumber \\
&  & \, -\frac{1}{2}
\sum_{j=1}^{N}
\int {\rm d}^4 x
\left(
\eta^{\mu \nu} \nabla_\mu  \Psi_j \nabla_\nu \Psi_j
+
{g^2} \Phi^2 \Psi_j^2
\right)
\; .
\label{classical-action}
\end{eqnarray}
One can divide the fields into classical and quantum parts by writing
$\Phi = \phi_c + \hat \phi$ and $\Psi_j = \hat \psi$ with $\langle
\Phi \rangle = \phi_c$ and $\langle \Psi_j \rangle = 0$.  Then if the
inflaton field is rescaled so that $\phi_c \rightarrow \sqrt{N}\phi_c$
it is possible to carry out a large $N$ expansion\footnote{The
  expansion here is similar in nature to one that was carried out for
  Quantum Electrodynamics with $N$ fermion fields in
  Ref.~\cite{largeN}.} using the closed time path
formalism~\cite{ctp}.  The result to leading order is that the system
is equivalent to a two field system with the classical inflaton field
$\phi_c$ coupled to a single quantized field $\hat \psi$, henceforth
referred to simply as $\phi$ and $\psi$, respectively.  The equations
of motion for these fields are
\begin{subequations}
\begin{eqnarray}
\left(
-\Box + m^2 + g^2 \langle \psi^2\rangle_{\rm B} \right) \phi&=&0
\; ,
\label{motion-phi-1}
\\
\left(
-\Box + g^2 \phi^2 \right) \psi&=&0
\; ,
\label{motion-psi-1}
\end{eqnarray}
\end{subequations}
where $\langle \psi^2\rangle_{\rm B}$ is the bare (unrenormalized)
expectation value of $\psi^2$.  Restricting to the case of a
homogeneous classical inflaton field in Minkowski spacetime, the
equation of motion simplifies to
\begin{eqnarray}
\ddot \phi(t) +( m^2 + g^2 \langle  \psi^2 \rangle_{\rm B} ) \phi(t) &=&0
\; .
\label{motion-phi-2}
\end{eqnarray}
Equation~\eqref{motion-psi-1} is separable and the quantum field
$\psi(x)$ can then be expanded as
\begin{equation}
\psi({\bf x},t) =
\int \frac{{\rm d}^3 {\bf k}}{(2 \pi)^{3/2}}
\left[ a_{\bf k} \, e^{i {\bf k} \cdot {\bf x}} \, f_k (t)
+ a_{\bf k}^\dagger \, e^{-i {\bf k} \cdot {\bf x}}\,f^*_k (t) \right]
\; .
\label{psi-mode}
\end{equation}
The time dependent modes, $f_k(t)$, satisfy the following ordinary
differential equation
\begin{eqnarray}
\ddot f_k  (t)+ [{k^2}+ g^2 \phi^2(t)] f_k(t)&=&0
\; ,
\label{mode-eq-1}
\end{eqnarray}
and the Wronskian (normalization) condition
\begin{eqnarray}
f_k \frac{\rm d}{\rm d t} f_k^*
-
f_k^* \frac{\rm d}{\rm d t} f_k &=& i
\; .
\label{wronskian-eq-1}
\end{eqnarray}

The WKB approximation for these modes is useful both for
renormalization and fixing the state of the field.  It is
obtained by writing
\begin{equation}
f_k (t) =
\frac{1}{\sqrt{2 W_k (t)}}
\exp\left[
-i \int_{t_0}^t  {\rm d} t' \; W_k(t')
\right]
\; .
\label{eq:wkbmodes}
\end{equation}
Substitution of Eq.~\eqref{eq:wkbmodes} into Eq.~(\ref{mode-eq-1}) gives
\begin{equation}
 W_k^2 = k^2 + g^2 \phi^2 - \frac{1}{2}
\left(\frac{\ddot{W_k}}{W_k}
- \frac{3}{2} \frac{\dot{W_k}^2}{W_k^2} \right)
\; .
 \label{W2}
 \end{equation}
One solves this equation by iteration.  Upon each iteration one
obtains a WKB approximation which is higher by two orders than the previous one.
Note that the order depends both upon the number of time derivatives
and the power of $g$.
The zeroth
order approximation is just $W_k^{(0)} = k$, while the second order
approximation is
\begin{equation}
W_k^{(2)} = k + \frac{ g^2 \phi^2}{2 k}
\; ,
\label{Wsecond}
\end{equation}
and the fourth order one is
\begin{equation}
W_k^{(4)} = k + \frac{ g^2 \phi^2}{2 k} -\frac{g^4 \phi^4}{8 k^3}
-\frac{g^2}{4 k^3} (\phi \ddot{\phi} + \dot{\phi}^2)
\; .
\label{Wfourth}
\end{equation}

Eq.~(\ref{motion-phi-2}) for the inflaton field involves the quantity
$\langle \psi^2 \rangle_{\rm B}$. Using~\eqref{psi-mode} one finds
\begin{equation}
\langle  \psi^2 \rangle_{\rm B}
= \frac{1}{2 \pi^2} \int_{0}^{+\infty} {\rm d} k \; k^2
\; \left| f_k(t) \right|^2
\; .
\label{bare-psi-squared}
\end{equation}
This quantity is divergent and must be regularized. We use the method
of adiabatic regularization~\cite{parker,zel-star,par-ful,bunch}
which, for free scalar fields in Robertson-Walker spacetimes has been
shown to be equivalent to the covariant scheme of point
splitting~\cite{birrell,pa}.

In adiabatic regularization the renormalization counterterms are
obtained by using a WKB approximation for the modes of the quantized
field~\cite{par-ful,bunch,birrell,pa,ae,amr}. One works with a massive field and
then takes the zero mass limit at the end of the
calculation.  Expressions for the renormalized values of
both $\langle \psi^2 \rangle$ and the expectation value of the
energy-momentum tensor $\langle T_{\mu \nu} \rangle$ for a similar
system to the one we are using\footnote{In Ref.~\cite{amr}, if the
  scale factor, $a(t)$, is set equal to one, the coupling of the
  inflaton field to the scalar curvature is set to zero, the coupling
  constant $\lambda$ is set equal to $2 g^2$, the mass of the quantum
  field is set to zero, and the $\lambda \phi^3/3!$ term in the
  equation of motion for the $\phi$ field is dropped, then the two
  systems are equivalent.}  have been obtained in Ref.~\cite{amr}.
Some details of the renormalization procedure are given in the
Appendix.  The result for our system is
\begin{subequations}
\begin{eqnarray}
\langle  \psi^2 \rangle_{\rm ren}
&=& \frac{1}{2 \pi^2} \int_{0}^{\epsilon} {\rm d} k \; k^2
\;\left( \left| f_k(t) \right|^2 - \frac{1}{2k} \right) \nonumber \\
& & \nonumber \\ &  &  \, +
\frac{1}{2 \pi^2} \int_{\epsilon}^{+\infty} {\rm d} k \; k^2
\;\left( \left| f_k(t) \right|^2 - \frac{1}{2k}
+ \frac{g^2 \phi^2}{4k^3} \right) + \langle \psi^2 \rangle_{\rm an}
\; ,
\label{psi2ren-a}
\\   &  & \nonumber \\
 \langle \psi^2 \rangle_{\rm an} &=& - \frac{g^2 \phi^2}{8 \pi^2}
\left[1 - \log \left( \frac{2 \epsilon }{M}\right)
\right]
\; ,
\end{eqnarray}
\label{psi2ren}
\end{subequations}
where $\epsilon$ is a lower limit cutoff that is placed in the
integrals that are infrared divergent and $M$ is an arbitrary
parameter with dimensions of mass which typically appears when the
massless limit is taken~\cite{ae,amr}.  In principle the value of $M$
should be fixed by observations.  For simplicity, we set it equal to
the mass of the inflaton field, $m$, in the calculations below. Note
that the value of $\langle \psi^2 \rangle_{\rm ren}$ is actually
independent of the value of the infrared cutoff $\epsilon$.

It is useful for both the analytic and numerical calculations to scale
the mass of the inflaton field, $m$, out of the problem.  This can be
done by defining new dimensionless variables as follows
\begin{subequations}
\begin{eqnarray}
\bar t &=& m t
\; ,
\label{t-scaling}
\\
\bar \phi &=& \frac{\phi}{m}
\; ,
\label{phi-scaling}
\\
\bar k &=& \frac{k}{m}
\; ,
\label{k-scaling}
\\
\bar f_k &=& \sqrt{m} f_k
\; ,
\label{fk-scaling}
\\
\bar M &=& \frac{M}{m}
\;, \\
\bar \epsilon &=& \frac{\epsilon}{m} \;.
\label{epsilon-scaling}
\end{eqnarray}
\label{scaled-equations}
\end{subequations}
If one substitutes $\langle \psi^2 \rangle_{\rm ren}$ for $\langle \psi^2 \rangle_{B}$ in
Eq.~\eqref{motion-phi-2}, writes
Eqs.~\eqref{motion-phi-2},~\eqref{mode-eq-1},~\eqref{wronskian-eq-1},
and~\eqref{psi2ren} in terms of the dimensionless variables, rescales $\langle \psi^2 \rangle_{\rm ren}$ so that
$\langle \psi^2 \rangle_{\rm ren} \rightarrow m^{-2} \langle \psi^2 \rangle_{\rm ren}$, and then
drops the bars, the equation for the inflaton field
[Eq.~\eqref{motion-phi-2}] becomes
\begin{eqnarray}
\ddot \phi(t) +  [1 + g^2 \langle  \psi^2 \rangle_{\rm ren}] \phi(t) &=&0
\; ,
\label{motion-phi-scaled}
\end{eqnarray}
while Eqs.~\eqref{mode-eq-1},~\eqref{wronskian-eq-1}
and~\eqref{psi2ren} remain the same.  These four equations in terms of
the dimensionless variables are the ones that are solved in the
following sections.


\section{Background Field Approximation}
\label{parametric}

In the background field approximation the wave equation for the
inflaton field is solved without taking into account the backreaction
effects of the quantum fields on the inflaton field.  However, the
effect of the inflaton field on the mode equations for the quantum
fields is taken into account.  For our model the wave equation for the
inflaton field~\eqref{motion-phi-scaled} then is a simple harmonic
oscillator equation, and the mode equation Eq.~\eqref{mode-eq-1} is a
Mathieu equation.

It is well known that for the Mathieu equation there are regions in
parameter space for which there are solutions which grow exponentially
due to a process called parametric amplification.
For the mode equation~\eqref{mode-eq-1} the last term is proportional
to $g^2 \phi_0^2$, with $\phi_0$ the amplitude of the oscillations of
the inflaton field.  Thus, for a given $g^2 \phi_0^2$, it is the
parameter $k$ which determines which modes undergo parametric
amplification.  There are bands of values of $k$ for which this occurs
and we shall call them instability bands.  The instability band which
contains the smallest values of $k$ will be called the first band.

For a given value of $g^2 \phi_0^2$ there is an infinite number of instability
bands some of which contain modes with arbitrarily large values of
$k$.  In principle, this could lead to divergences in quantities such
as $\langle \psi^2 \rangle_{\rm ren}$ and the expectation value of the
energy-momentum tensor, $\langle T_{\mu \nu} \rangle_{\rm ren}$.  The
reason is that, as can be seen from Eq.~\eqref{psi2ren-a}, the
renormalization counterterms do not undergo any exponential growth
while bands of modes with arbitrarily large values of $k$ do.  One
might also be concerned that, even if $\langle \psi^2 \rangle_{\rm
  ren}$ and $\langle T_{\mu \nu} \rangle_{\rm ren}$ are are finite, in
order to compute them one must take into account modes in instability
bands with arbitrarily large values of $k$.  This is clearly not
possible for purely numerical computations, which must necessarily
include only a finite number of modes and for which there must be an
ultraviolet cutoff.  In this section we use known properties of
solutions to the Mathieu equation to show that both $\langle \psi^2
\rangle_{\rm ren}$ and $\langle T_{\mu \nu} \rangle_{\rm ren}$ are
finite at any finite time.  We also address the question of how to
make sure that the contributions of all of the instability bands which
contribute significantly to $\langle \psi^2 \rangle_{\rm ren}$ and
$\langle T_{\mu \nu} \rangle_{\rm ren}$ are included in numerical
calculations of these quantities.


\subsection{The Mathieu Equation}
\label{mathieu-section}

In the background field approximation, without loss of generality (due
to time translation invariance), we can take the solution to
Eq.~\eqref{motion-phi-scaled} to be
\begin{equation}
\phi (t) = \phi_0 \cos (t)
\; .
\label{eq:phicos}
\end{equation}
The mode equation is then
\begin{eqnarray}
\ddot f_k (t) + [{k^2}+ g^2 \phi_0^2 \cos^2(t) ] f_k (t) &=&0
\; .
\label{par-psi-mathieu-1}
\end{eqnarray}
Using the identity
\begin{eqnarray}
\cos^2(t)&=& \frac{1}{2} [1+\cos(2t)]
\; ,
\end{eqnarray}
the mode equation can be put into a standard form for the Mathieu
equation~\cite{mathieu-book}
\begin{eqnarray}
\frac{{\rm d}^2 f_k(t)}{{\rm d}t^2}  +
[a - 2q  \cos(2t)] f_k (t)&=&0
\; ,
\label{mathieu-0}
\end{eqnarray}
with
\begin{subequations}
\begin{eqnarray}
a &=& k^2 + \frac{1}{2} g^2 \phi_0^2
\; ,
\label{aeq}
\\
q &=& - \frac{1}{4} g^2 \phi_0^2
\; .
\label{qeq}
\end{eqnarray}
\label{aqeq}
\end{subequations}

Floquet's theorem applies to the Mathieu
equation~\cite{mathews-walker}. It implies the existence of two
solutions of Eq.~(\ref{mathieu-0}) of the form:
\begin{eqnarray}
f_1(t)&=& e^{\mu t} h_1(t)
\; ,
\label{f1}
\\
f_2(t)&=& e^{-\mu t} h_2(t)
\; ,
\label{f2}
\end{eqnarray}
where $h_1 (t)$ and $h_2 (t)$ are periodic functions of period $2 \pi$
and $\mu$ is, in general, a complex number~\cite{mathews-walker}.  If
$\mu$ has a non-vanishing real part, it is easy to see that one
solution will grow exponentially at late times.

Our goal is to find which ranges of values of $k$ lead to exponentially
growing modes.  This can be accomplished by first fixing the value of
$q$ and then finding the values of the parameter $a$ for which the
Mathieu equation has solutions with $\mu = 0$.  Those values of $a$
which correspond to even solutions are traditionally labeled $a_r$ and
those which correspond to odd solutions are labeled $b_r$, with $r$ a
positive integer.  The pattern of instability and stability regions in
the $(a,q)$ plane is symmetric about $q=0$.  Thus, for simplicity we
shall consider $q>0$ in what follows, even though $q < 0$ for the
equations we are concerned with [see Eq.~\eqref{qeq}]. For $q >0$, it
is found that $\mu^2 >0$ for values of $a$ which are between $a_r$ and
$b_r$ for the same value of $r$~\cite{mathieu-book}.  Further one has
the relation $a_0 < b_1 < a_1 < b_2 < \dots \; $.

For the purpose of assessing the contribution of unstable modes with
large values of $k$ to Eq.~\eqref{psi2ren}, it is necessary to know
both the range of values of $k$ in the unstable bands and the maximum
value of $\mu$ in a band containing large values of $k$.  The ranges
of $k$ in the large $r$ limit are given by the
relation~\cite{abram-stegun}
\begin{equation}
a_r - b_r = O\left(\frac{q^r}{r^{r-1}} \right)
\; .
\label{width}
\end{equation}
Thus, for fixed $q$, the width of the unstable bands in the $(a,q)$
plane becomes arbitrarily small for bands with large values of $r$ and
therefore becomes arbitrarily small for large values of $k$.  This
means that for unstable modes at large values of $k$ to contribute
significantly to the integrals in Eq.~\eqref{psi2ren}, it would be
necessary that these modes grow much faster than the unstable modes in
bands with smaller values of $k$, which are wider.

An approximation to the value of $\mu$ in unstable bands
is~\cite{mathieu-book}
\begin{eqnarray}
\mu^2_r
&=&
\frac{(a_r+b_r-2a-4r^2)}{2}
\pm
\frac{\sqrt{a_r^2-2a_r b_r-8r^2(a_r+b_r-2a)+b_r^2+16r^4}}{2}
\; ,
\label{mu-1}
\end{eqnarray}
where the boundaries of the unstable band of index $r$ are given by
the pair of eigenvalues $b_r$ and $a_r$ with $a_r>b_r$.  Some details
of the derivation of this approximation can be found in Ref.~\cite{david-thesis}.  Note
that there are two possible solutions for $\mu_r$ for fixed values of
$r$ and $a$.  Given that for large $r$ the instability bands become
narrower, and that $b_r \le a \le a_r$, it is easy to show that for
large $r$ only the plus sign gives $\mu_r^2 > 0$.  It is also easy to
show that
\begin{equation}
|2a-a_r-b_r| \ll 4r^2
\; ,
\label{condition-mu}
\end{equation}
and thus that
\begin{equation}
\mu_r
\simeq \frac{[(a_r-a)(a-b_r)]^\frac{1}{2}}{{2r}}
\; .
\label{approx-mu-2}
\end{equation}
For this approximation the maximum value of $\mu_r$ is at the midpoint
between $b_r$ and $a_r$.
Since, for large $r$ the width of the band, $a_r - b_r$, gets very
small [see Eq.~\eqref{width}], it is clear that the maximum value of
$\mu$ in a band is smaller for bands with very large values of $k$
than for bands with smaller values of $k$.

The relative contribution of unstable modes to the integral in
Eq.~\eqref{psi2ren} at a given time $t$ depends in part on how much
they have grown compared to unstable modes in lower bands.  It is
clear from the above analysis that modes in higher bands grow
significantly slower than those in lower bands.  Also for a given time
$t$, there will be some value, $k = K$, for which unstable bands with
larger values of $k$ have not grown much at all.  This, coupled with
the fact that the widths of the unstable bands get narrower at larger
values of $k$, means that at any given time $t$ the modes in bands
with $k > K$ will not contribute significantly to the mode integral in
Eq.~\eqref{psi2ren}.  Thus, $\langle \psi^2 \rangle_{\rm ren}$ and
$\langle T_{\mu \nu} \rangle_{\rm ren}$ are finite.

{From} the above analysis it should be clear that in numerical
computations it is possible to obtain a good approximation to $\langle
\psi^2 \rangle_{\rm ren}$ and $\langle T_{\mu \nu} \rangle_{\rm ren}$
by putting in a cutoff at some large value of $k$. Since the location
of the bands at small values of $k$ can be determined using standard
techniques~\cite{mathieu-book}, then one simply needs to use a high
enough density of modes so that those bands with smaller values of $k$
are adequately covered.

The maximum value of $\mu$ occurs for the first unstable band which
has the smallest values of $k$.  Once the modes in this band have
become sufficiently large, they make the major contribution to
$\langle \psi^2 \rangle_{\rm ren}$ and $\langle T_{\mu \nu}
\rangle_{\rm ren}$.


\section{Backreaction on the Inflaton Field}
\label{results}


\subsection{Initial Conditions}
\label{results-initial}

A Fortran program was written to simultaneously solve the mode
equations for the $\psi$ field, compute the quantity $\langle \psi^2
\rangle_{\rm ren}$, and solve the equation for the inflaton field
$\phi$ in terms of the dimensionless
variables~\eqref{scaled-equations}.  For this program to run, initial
values must be given for $\phi$ and its first time derivative, as well
as each mode function $f_k$ and its first time derivative.  Since the
inflaton field is in its rapid oscillation phase, what is important is
the initial amplitude of its oscillations.  This is most easily
obtained by starting with $\phi(0) = \phi_0$ and $\dot{\phi}(0) = 0$.

The initial values for the mode functions of the $\psi$ field are
determined by the state of the field.  As the quantum field is coupled
to the inflaton field, there is no preferred vacuum state.  Instead
one can use an adiabatic vacuum state.  Such states have been
discussed in detail for cosmological spacetimes~\cite{b-d}.  They are based on the WKB
approximation for the modes.  It is necessary to choose a state that
is of adiabatic order two or higher for the renormalized value of
$\langle \psi^2 \rangle$ to be finite.  For the renormalized
energy-momentum tensor $\langle T_{\mu \nu}\rangle $ to be finite it
is necessary to have at least a fourth order adiabatic state.  For
such a state, in the large $k$ limit,
\begin{equation}
|f_k|^2 \rightarrow \frac{1}{2k} - \frac{g^2 \phi^2}{4 k^3} +
\frac{3 g^4 \phi^4}{16 k^5} + \frac{g^2 (\dot{\phi}^2
          + \phi \ddot{\phi})}{8 k^5} + O(k^{-6})
\; .
\end{equation}

Although the large $k$ behavior of an adiabatic state of a given order
is constrained by requiring that renormalization of the
energy-momentum tensor be possible, there are no such constraints on
modes with smaller values of $k$.  Thus, one can construct fourth
order adiabatic states that would give virtually any value for
$\langle \psi^2 \rangle_{\rm ren}$ or $\langle T_{\mu \nu}
\rangle_{\rm ren}$ at some initial time $t$.  However, if the value of
$g^2 \langle \psi^2 \rangle_{\rm ren}$ is too large at the beginning
of preheating, then backreaction effects will be important
immediately.  In fact, if $g^2 \langle \psi^2 \rangle_{\rm ren} \gg
1$, then backreaction effects are so strong initially that parametric
amplification is unlikely to occur.  If $g^2 \langle \psi^2
\rangle_{\rm ren} \sim 1$, backreaction effects will be important, but
it is possible that parametric amplification might still occur.  On
the other hand if $g^2 \langle \psi^2 \rangle_{\rm ren} \ll 1$ at the
onset of preheating, then backreaction effects are not important
initially and parametric amplification will occur. Since our goal is
to study the preheating process in detail, we restrict attention to
this latter case.

A natural way to construct a fourth order adiabatic state is by using
a fourth order WKB approximation to fix the initial values of the
modes, $f_k$, and their derivatives, $\dot f_k$, for all values of
$k$.  For the numerical calculations done in this paper the state used
was an adiabatic vacuum state with
\begin{subequations}
\begin{eqnarray}
f_k(t=0) &=& \sqrt{\frac{Y}{2}}
\; ,
\label{initial-fk}
\\
\dot{f}_k(t=0) &=& -\frac{i}{\sqrt{2Y}}
\; ,
\label{initial-fkdot}
\\
Y &=& \frac{1}{\Omega_0}
+ \frac{g^2 \phi_0 \ddot{\phi}_0}{4 \Omega_0^5}
\; ,
\label{y}
\\
\Omega_0 &=& \sqrt{k^2 + g^2 \phi_0^2}
\; .
\label{omegazero}
\end{eqnarray}
\label{fwkb}
\end{subequations}
It is easy to show that this is a fourth order adiabatic state by
recalling that $\dot{\phi}(0) = 0$.  The initial values have been
chosen in part to make it possible to analytically compute the initial
value of $\langle \psi^2 \rangle_{\rm ren}$, and in part so that the
Wronskian condition~\eqref{wronskian-eq-1} is satisfied exactly.  For
the fourth order adiabatic state that we are using, at the initial
time $t=0$
\begin{eqnarray}
g^2 \langle \psi^2 (t=0) \rangle_{\rm ren} =
\frac{g^2  \ddot{\phi}_0}{48 \pi^2 \phi_0}
  - \frac{g^4 \phi_0^2}{16 \pi^2}
\left[ 1 - \log \left(\frac{g^2 \phi_0^2}{M^2} \right) \right]
\; .
\end{eqnarray}
Using this initial value for $g^2 \langle \psi^2 \rangle_{\rm ren}$
one finds that the initial value for $\ddot \phi (t)$ is
\begin{equation}
\ddot{\phi} (t=0)=
\frac{1}{1+ \frac{g^2}{48 \pi^2}} \,
\left[-1 + \frac{g^4 \phi_0^2}{16 \pi^2}
\left(1 - \log\left(\frac{g^2 \phi_0^2}{M^2}
\right)\right) \right]\phi_0 \;.
\label{phipp}
\end{equation}
Thus, whether backreaction effects are important initially depends on
the values of both $\phi_0$ and $g$.  For the values of $g$ considered
here, $\frac{g^2}{48 \pi^2} \ll 1$.  {From} Eq.~\eqref{phipp} it can
be seen that backreaction effects for the states we are considering
are likely to be important if
\begin{equation}
g^2 (g^2 \phi_0^2) \stackrel{>}{_\sim} {16 \pi^2}
\; .
\label{g2g2phi2}
\end{equation}
As an example, for realistic models of chaotic inflation
KLS~\cite{KLS} chose
\begin{eqnarray}
m &=& 10^{-6} M_p
\; ,
\nonumber \\
\phi_0 &=& \frac{M_p}{5 m}
\; .
\end{eqnarray}
The values they used for $g$ range from $ \sim 10^{-4}$ to $10^{-1}$.
For $g = 10^{-4}$, $g^2 \phi_0^2 = 400$, while for $g = 10^{-1}$, $g^2
\phi_0^2 = 4 \times 10^8$.  Applying the criterion~\eqref{g2g2phi2}
one finds that our fourth order adiabatic states result in significant
backreaction initially for these values of $m$ and $\phi_0$ if $ g
\stackrel{>}{_\sim} 8 \times 10^{-3}$.


\subsection{Parametric Amplification With Backreaction}
\label{3backreaction}

If the initial conditions are such that $g^2 \langle \psi^2
\rangle_{\rm ren} \ll 1$, then examination of
Eq.~\eqref{motion-phi-scaled} shows that there is no significant
backreaction at early times and the solution of the inflaton field
will be approximately equal to Eq.~\eqref{eq:phicos}.  When, primarily
through parametric amplification of certain modes, $g^2 \langle \psi^2
\rangle_{\rm ren} \sim 1$, then one expects backreaction effects to be
important and to make the amplitude of the oscillations of the
inflaton field, $\phi$, decrease.  It has been observed in the
numerical computations that $\langle \psi^2 \rangle_{\rm ren}$
oscillates about an average value which changes in time.  The
effective mass of the inflaton field can be defined in terms of this
average value:
\begin{equation}
m_{\rm eff}^2  = 1 + g^2 \overline{\langle \psi^2 \rangle}_{\rm ren}
\; .
\label{meff}
\end{equation}
It is clear that an increase in $m_{\rm eff}^2$ results in an increase
of the frequency of oscillations of $\phi$.  This should happen slowly
at first, and so one would expect the basic character of the mode
equation~\eqref{mode-eq-1} to be a Mathieu equation, except that the
effective value of $q$ would change~\cite{KLS}. A zeroth order WKB
analysis of the solutions to Eq.~\eqref{motion-phi-scaled} shows that
the effective frequency of oscillations can be approximated by
\begin{equation}
\omega_{\rm eff} \approx \frac{1}{t_2 - t_1} \;
\int_{t_1}^{t_2} {\rm d} t \; \sqrt{1 + g^2 \langle \psi^2 \rangle_{\rm ren}}
\; ,
\label{omega-eff}
\end{equation}
with $t_1 < t < t_2$ for some time interval $t_2 - t_1$ equal to or
larger than an oscillation period of the inflaton field.  A derivation
similar to that at the beginning of Sec.~III then shows that
\begin{subequations}
\begin{eqnarray}
a_{\rm eff} & \approx & \frac{k^2}{\omega_{\rm eff}^2} + 2 q_{\rm eff}
\; ,
\label{aeff}
\\
q_{\rm eff} & \approx & \frac{g^2 A^2}{4 \omega_{\rm eff}^2}
\; ,
\label{qeff}
\end{eqnarray}
\end{subequations}
where $A$ is the amplitude of the inflaton field and we have defined
$q_{\rm eff}$ so that it is positive\footnote{The actual derivation
  results in a negative value for $q_{\rm eff}$ but the symmetry between
  positive and negative values of $q$ for the Mathieu equation allows
  us to simply consider positive values.}.  As the inflaton field
evolves $A$ will tend to slowly decrease and $\omega_{\rm eff}$ will
tend to slowly increase.  Thus, the resulting changes in the
instability bands should be small enough that there would be no
problem in having enough modes to accurately determine the primary
contributions to $\langle \psi^2 \rangle_{\rm ren}$.  Our numerical
work described below seems to bear this out.

Once the backreaction is significant, it is possible for the character
of the equation to change substantially.  If $\phi$ varies in a
periodic manner then it is still possible for parametric amplification
to occur due to Floquet's Theorem~\cite{mathews-walker}.  It would be
difficult in this more complicated situation to analyze the behavior
of the solutions to the extent that solutions to the Mathieu equation
have been analyzed.  However, it seems quite likely that, as for the
Mathieu equation, the most important instability bands will be at
relatively small values of $k$ and that if an instability band is
extremely narrow, then it probably will not have a large effect on
$\langle \psi^2 \rangle_{\rm ren}$ for a very long time.  It is likely
that interactions which are neglected in our model, will be
important on such long time scales~\cite{felder-kofman,tkachev,prokopec-roos,berges-serreau,pfkp,dfkpp}.


\subsection{Some Numerical Results}
\label{results-num}

\begin{figure}
\vskip -0.2in \hskip -0.4in
\includegraphics[angle=90,width=3.4in,clip]
{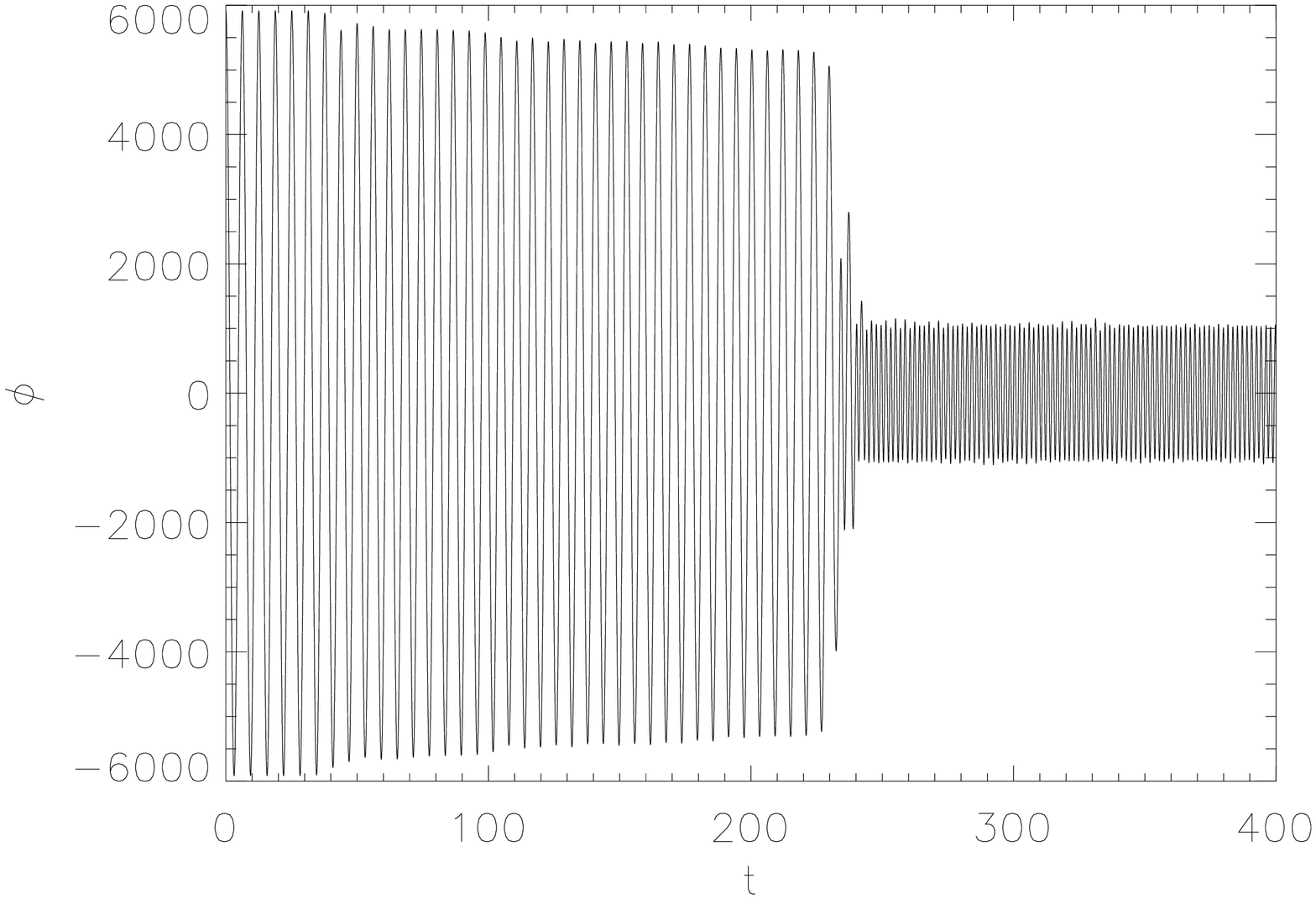}
\includegraphics[angle=90,width=3.4in,clip]
{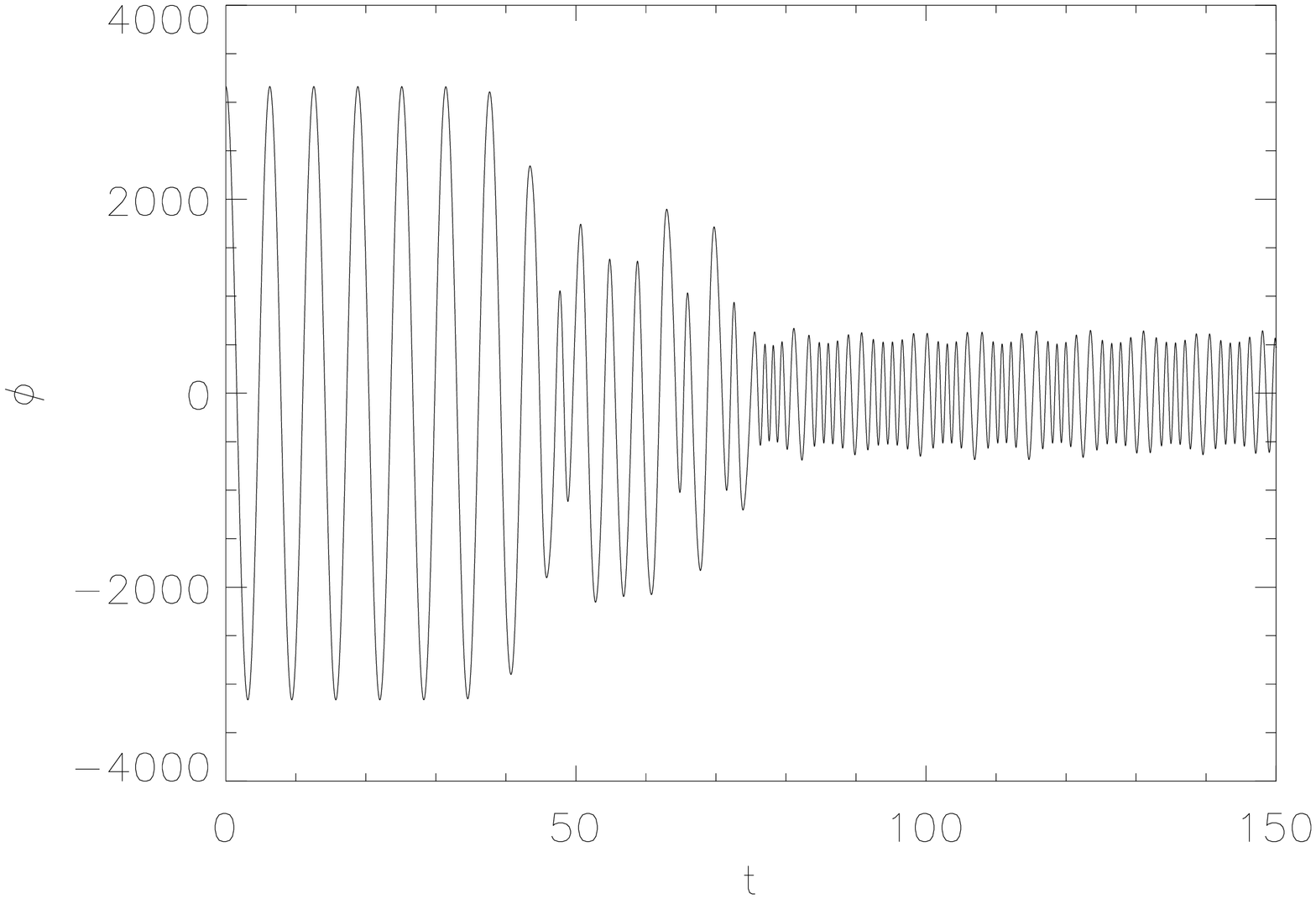}
\vskip -0.2in \hskip -0.4in
\includegraphics[angle=90,width=3.4in,clip]
{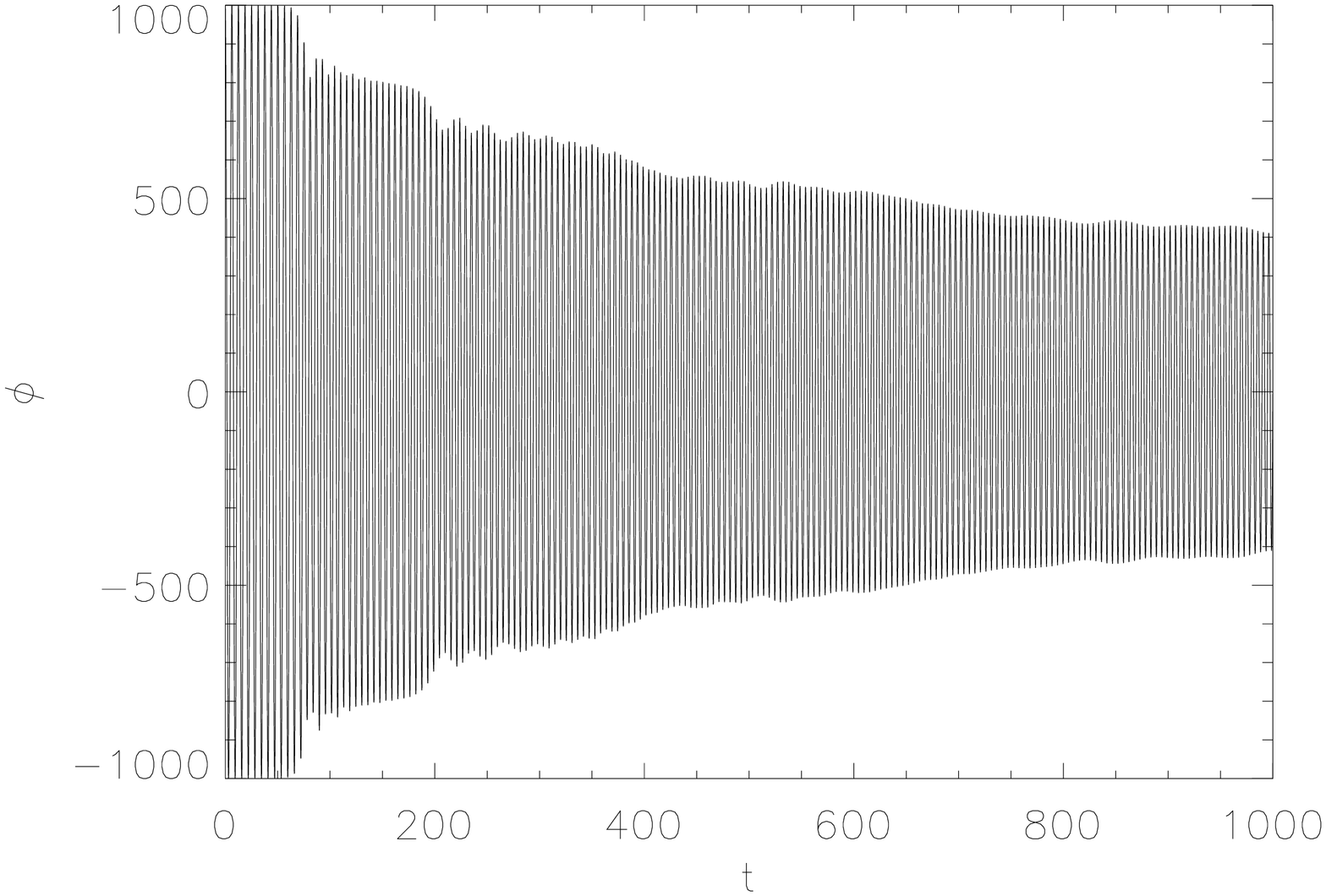}
\vskip -.2in
\caption{The evolution of the inflaton field for $g = 10^{-3}$.  From
  left to right the top plots are for $g^2 \phi_0^2 = 35$ and $10$.
  The bottom plot is for $g^2 \phi_0^2 = 1$.}
\label{fig1}
\end{figure}

\begin{figure}
\vskip -0.2in \hskip -0.4in
\includegraphics[angle=90,width=3.4in,clip]
{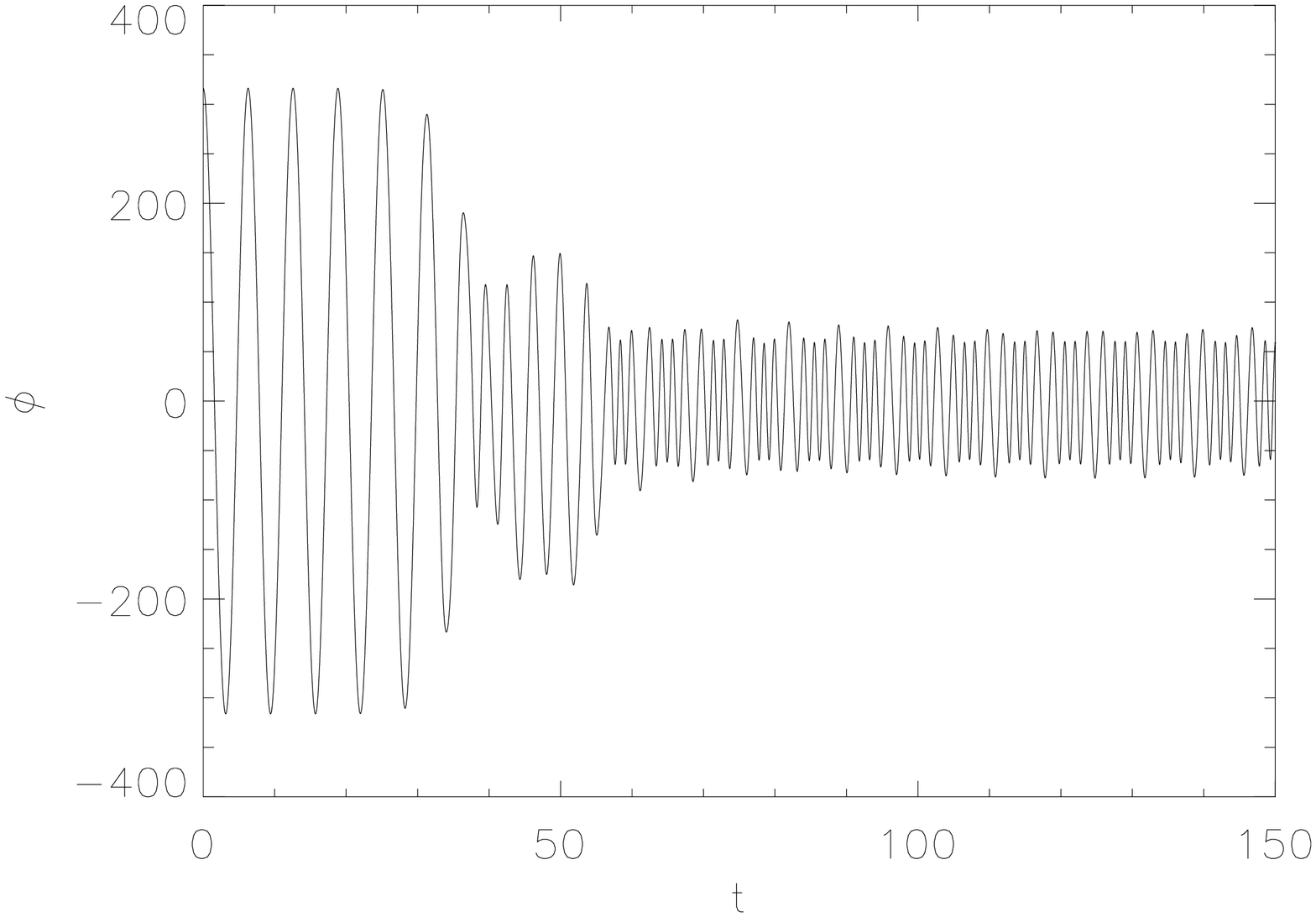}
\includegraphics[angle=90,width=3.4in,clip]
{fig_1_gphi2_10.ps}
\vskip -.2in \caption{The evolution of the inflaton field for $g^2 \phi_0^2 = 10$.
  The plot on the left is for
  for $g = 10^{-2}$ and the one on the right is for $10^{-3}$. }
\label{fig2}
\end{figure}

Since the mode equation~\eqref{mode-eq-1} depends on $g^2 \phi^2$ and
the equation of motion of the inflaton field~\eqref{motion-phi-scaled}
depends on $g^2 \langle \psi^2 \rangle_{\rm ren}$, one would expect
that the effects of the values of $g$ and $g^2 \phi_0^2$ decouple, at
least to some extent.  As can be seen in Figs.~\ref{fig1}
and~\ref{fig2} this is the case, with the quantity $g^2 \phi_0^2$
determining the type of evolution the inflaton field undergoes and $g$
having its strongest effect on the timescale of that evolution.

In Fig.~\ref{fig1} plots of solutions to the backreaction
equation~\eqref{motion-phi-scaled} for the case in which $g = 10^{-3}$
are shown, with $g^2 \phi_0^2$ ranging from $35$ to $1$. For the first
two plots most of the damping occurs over a very short period.  After
that, the field does not appear to damp significantly.  Instead, it
continues to oscillate but with a much higher frequency than
before. The envelope of its oscillations also oscillates but with a
much smaller frequency.  This behavior has been observed for all cases
investigated with $g^2 \phi_0^2 \stackrel{>}{_\sim} 2$.  Conversely,
for the plot on the bottom ($g^2 \phi_0^2 = 1$), it is clear that the
damping is much slower and occurs over a much longer period of time.
This behavior is a verification of the prediction by KLS~\cite{KLS}
that a period of rapid damping would occur for $g^2 \phi_0^2 \gg 1$
and that only relatively slow damping occurs for $g^2 \phi_0^2 \ll 1$.
A detailed analysis of the evolution
of the inflaton field is given in
subsection~\ref{results-details}.

In Fig.~\ref{fig2} plots of solutions to the backreaction equation for
the case in which $g^2 \phi_0^2 = 10$ and $g = 10^{-2}$ and $10^{-3}$
are shown.  The amount of damping does not appear to depend in any
strong way upon the value of the coupling constant $g$.  The amount of
time it takes for significant damping to begin is longer for the
smaller value of $g$.  This is easily explained by the fact that
initially $g^2 \langle \psi^2 \rangle_{\rm ren} \ll 1$, so that only
negligible damping of the inflaton field occurs at early times.
During this period the mode equations and thus the growth of $ \langle
\psi^2 \rangle_{\rm ren}$ is approximately independent of the value of
$g$.  Significant backreaction begins to occur when $g^2 \langle
\psi^2 \rangle_{\rm ren} \sim 1$ and this will obviously take longer
for smaller values of $g$, since $ \langle \psi^2 \rangle_{\rm ren}$
will have to grow larger before it can occur.

The natural explanation for what is happening in all of these cases is
that particle production occurs and the backreaction of the produced
particles on the inflaton field causes its amplitude to be damped.
However, if this picture is correct then one would expect that a
significant amount of energy would be transferred from the inflaton
field to the modes of the quantum field.  To check this, one can
compute the time evolution of the energy density of the inflaton
field.  Since the gravitational background is Minkowski spacetime, the
energy density is conserved as is shown explicitly in the Appendix.
As is also shown in the Appendix, the energy density can be broken
into two different contributions.  The first one is the energy density
the inflaton field would have if there was no interaction ($g = 0$).
It is given by
\begin{equation}
\rho_\phi = \frac{1}{2} \left( \dot{\phi}^2 + m^2 \phi^2 \right)
\; .
\label{rhophi}
\end{equation}
The other contribution is the energy density of the quantum field
shown in Eq.~\eqref{t00-quant}.  It contains terms that would be there
if there was no interaction along with terms that explicitly depend on
the interaction.  This split is useful because almost all of the
energy is initially in the inflaton field and thus, one can clearly
see how much has been transferred to the quantum field $\psi$ and to the
interaction between the two fields as time
goes on.

\begin{figure}
\vskip -0.2in \hskip -0.4in
\includegraphics[angle=90,width=3.4in,clip]
{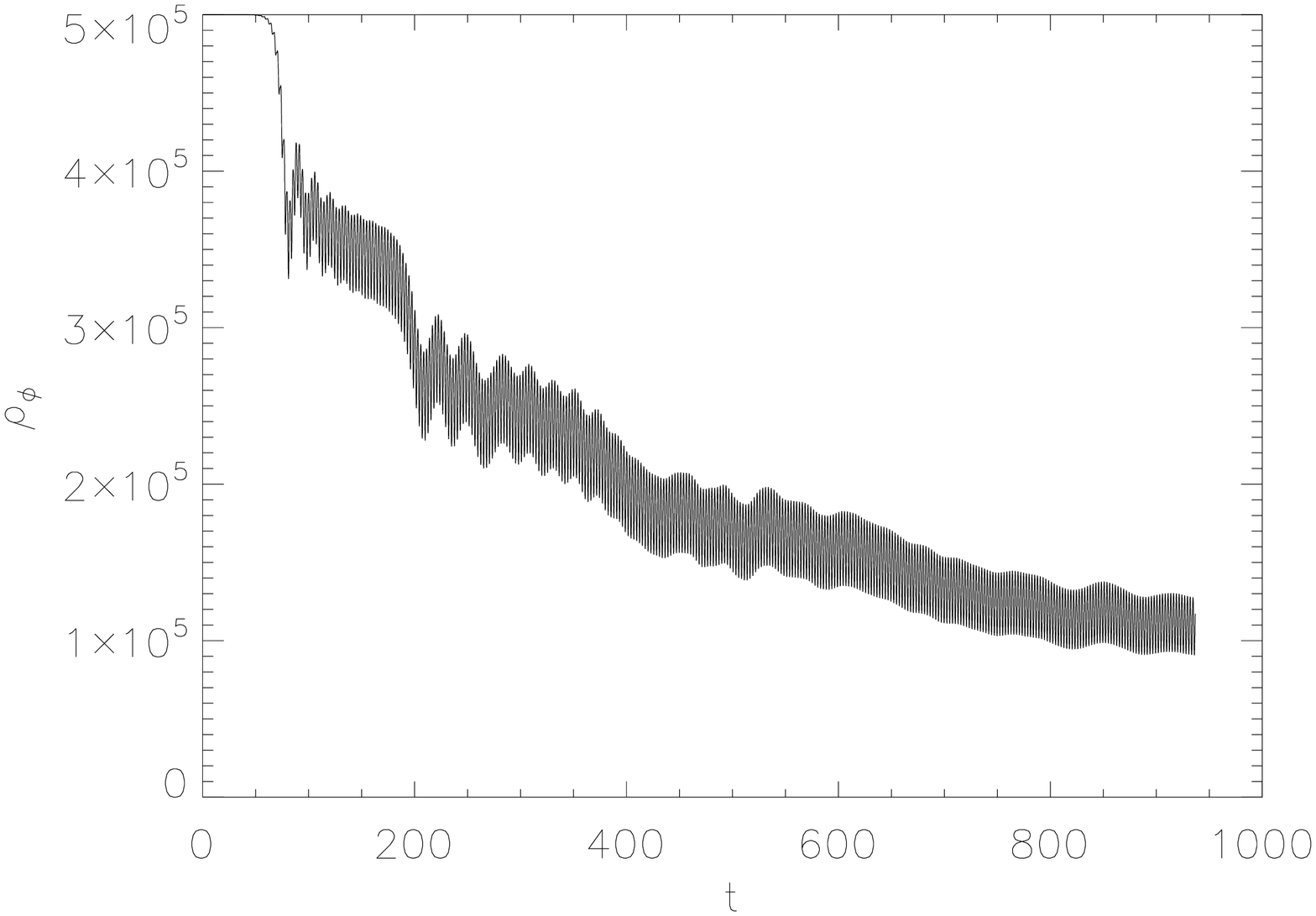}
\includegraphics[angle=90,width=3.4in,clip]
{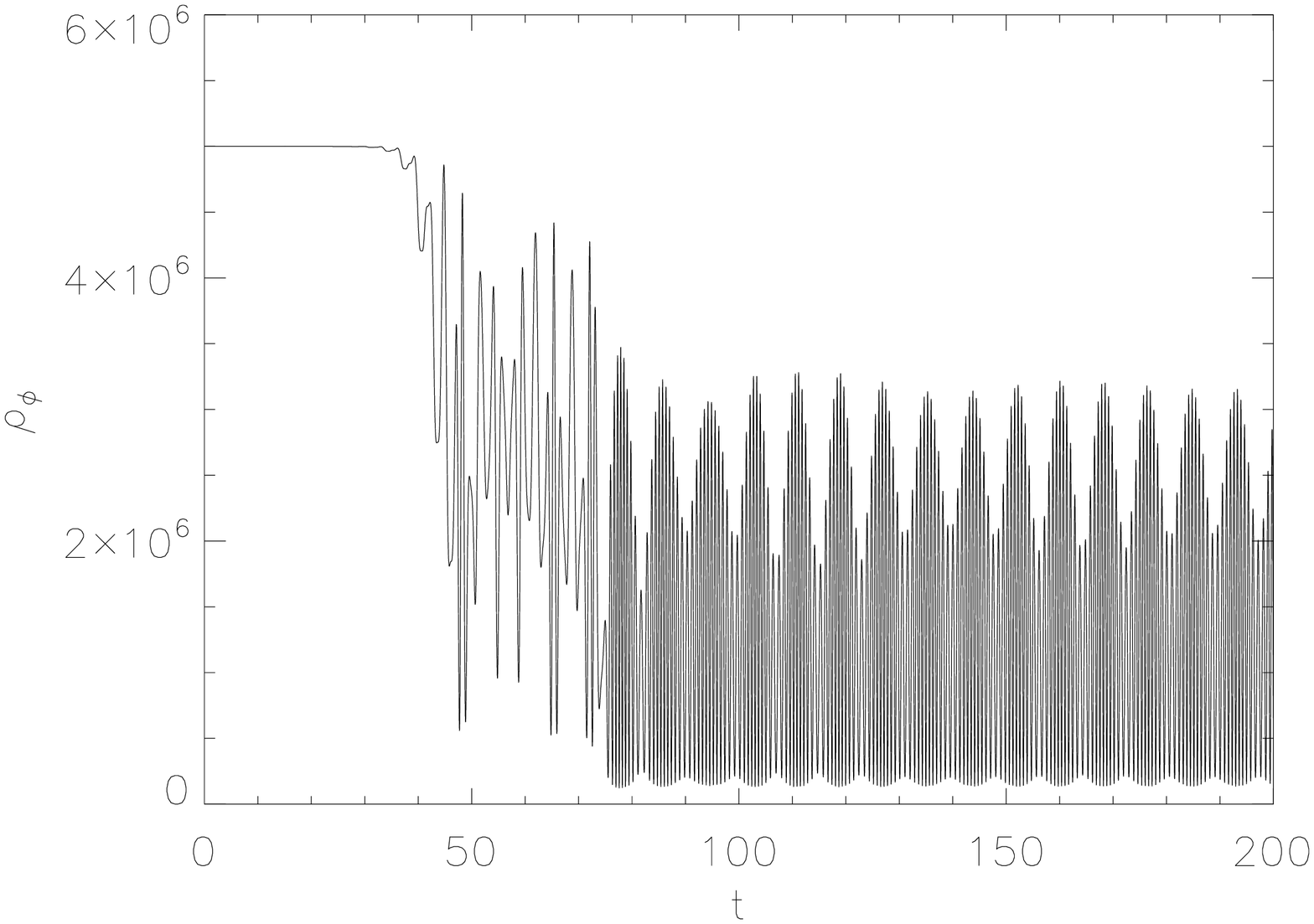}
\vskip -.2in \caption{The evolution of the energy density of the
  inflaton field, $\rho_\phi$.  Both plots are for $g = 10^{-3}$.
  The one on the left is for $g^2 \phi_0^2 =1$ and the one on the
  right is for $g^2 \phi_0^2 = 10$.}
\label{fig-energy}
\end{figure}

In Fig.~\ref{fig-energy} the left hand plot shows the evolution of the
energy density of the inflaton field for the case $g = 10^{-3}$ and
$g^2 \phi_0^2 = 1$.  Comparison with the bottom plot in
Fig.~\ref{fig1} shows that as the inflaton field is damped, a large
portion of its energy is permanently transferred away from it.

For all cases investigated in which rapid damping of the inflaton
field occurs, it was found that much less energy is permanently
transferred away from the inflaton field than one might expect, given
the amount by which its amplitude has been damped.  This effect was seen previously
in the classical lattice simulation of Prokopec and Roos~\cite{ prokopec-roos}.  An example is the
case $g = 10^{-3}$, $g^2 \phi_0^2 = 10$ for which the evolution of the
energy density of the inflaton field is shown in the plot on the right
panel of Fig.~\ref{fig-energy}.  Comparison with the top right plot of
Fig.~\ref{fig2} shows that after the rapid damping has finished the
inflaton field permanently loses some energy because the maximum of
its energy density is lower than before, but more than half of
the original energy is transferred back and forth between the inflaton
field, the quantum field, and the interaction between the two.  However, it
is important to note that in cases in which rapid damping occurs our results cannot
be trusted after the rapid damping phase has finished because, as previously mentioned, there is evidence
that the interactions we neglect become important~\cite{felder-kofman,tkachev,prokopec-roos,berges-serreau,pfkp,dfkpp}.

\begin{figure}
\vskip -0.2in \hskip -0.4in
\includegraphics[angle=90,width=3.4in,clip]
{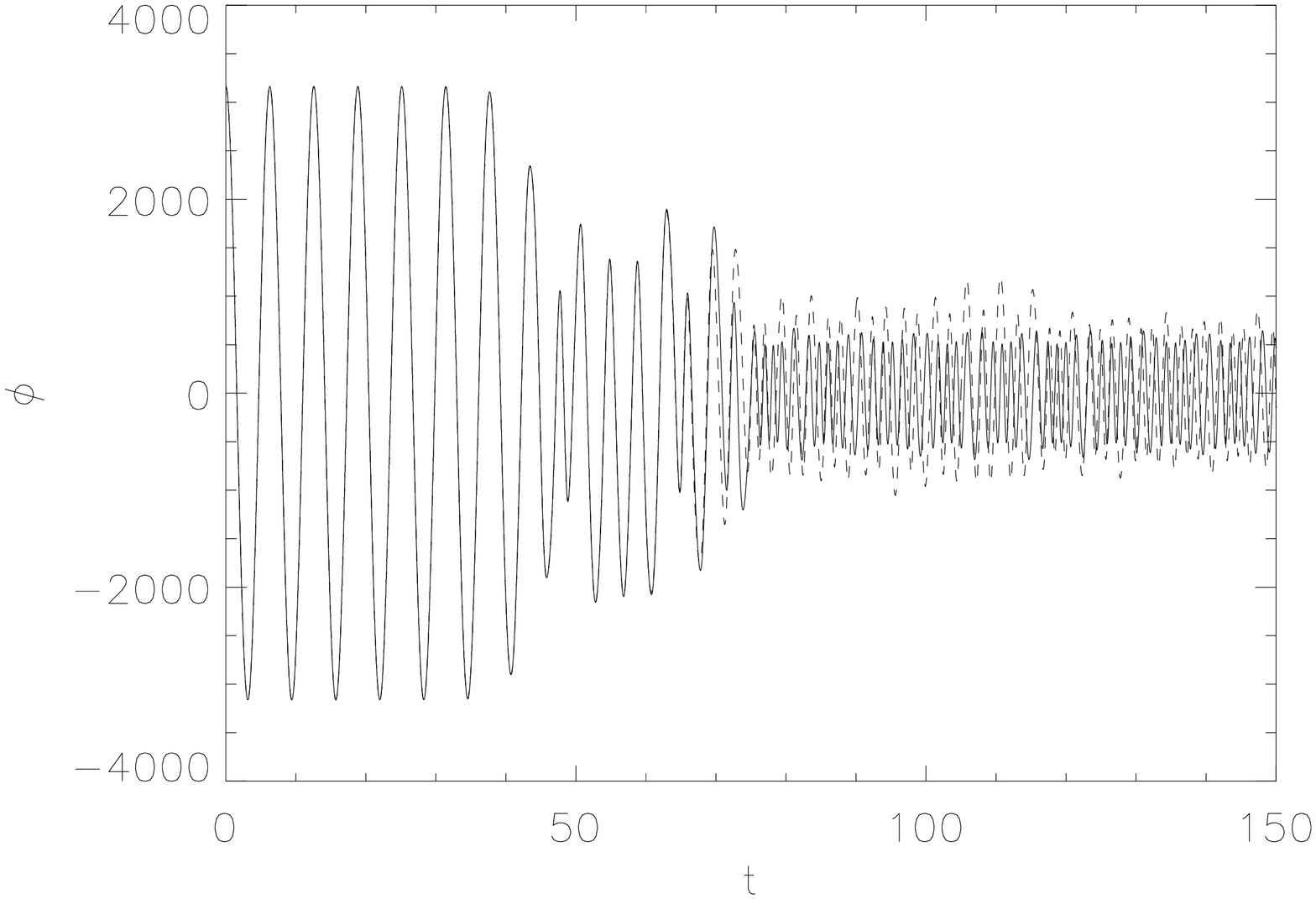}
\includegraphics[angle=90,width=3.4in,clip]
{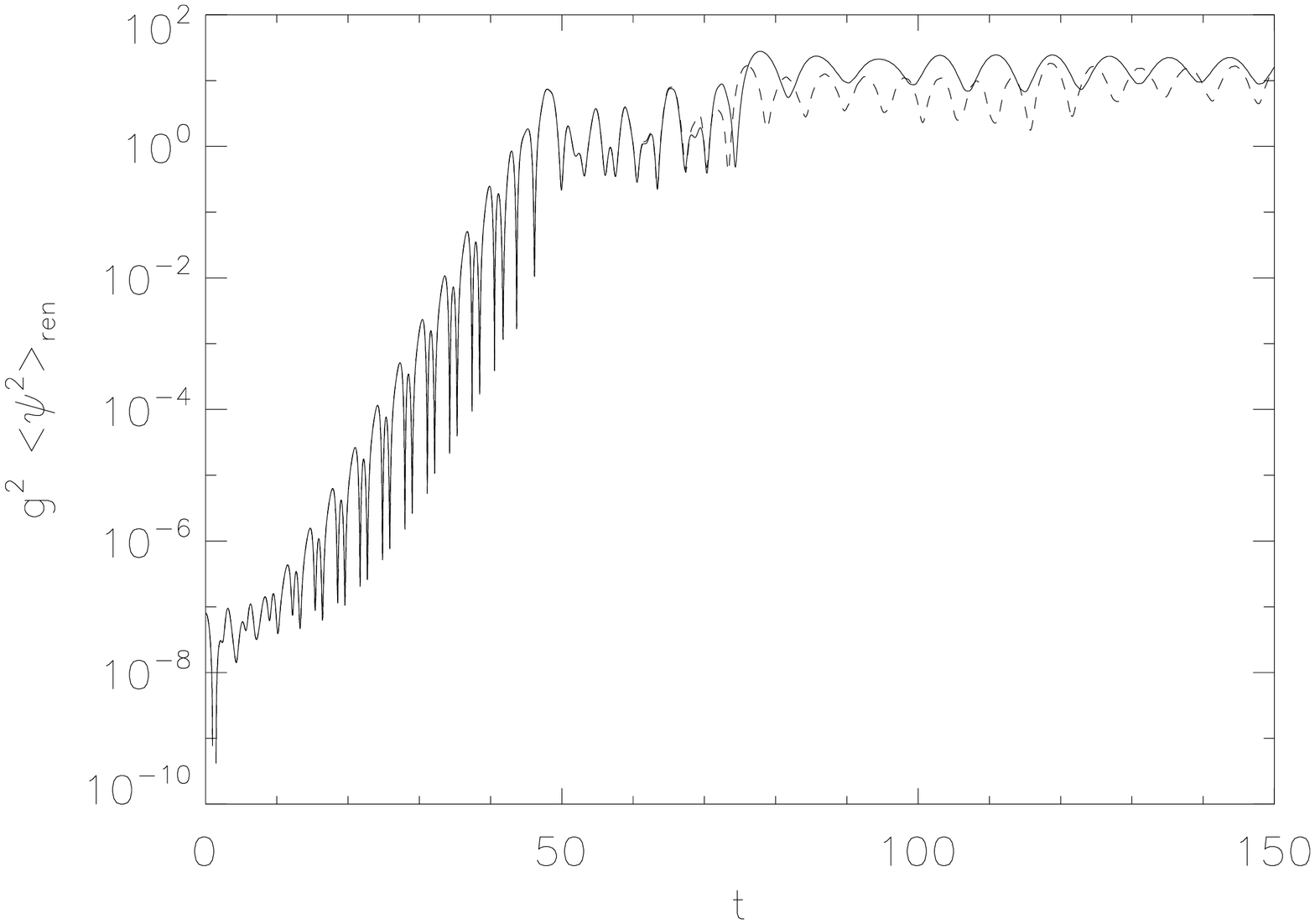}
\vskip -.2in \caption{This figure shows differences in $\phi$
and $g^2 \langle \psi^2 \rangle_{\rm ren}$ which occur for $g = 10^{-3}$ and
$g^2 \phi_0^2 = 10$ and $g^2 \phi_0^2 = 10(1+10^{-5})$.
In each plot the dashed curve corresponds to $g^2 \phi_0^2 = 10(1+10^{-5})$. }
\label{fig-comp}
\end{figure}

Another effect that was observed to occur in all cases in which the
inflaton field undergoes rapid damping, is an extreme sensitivity to
initial conditions.  If the initial conditions are changed by a small
amount then initially the evolution of the inflaton field and the
quantity $\langle \psi^2 \rangle_{\rm ren}$ are nearly the same as
before.  However, at some point the evolution of $\langle \psi^2
\rangle_{\rm ren}$ begins to change significantly and that is followed
by a significant change in the evolution of $\phi$.  An example is
shown in Fig.~\ref{fig-comp}.  This phenomenon is discussed in more
detail below.

\subsection{Detailed Analysis}
\label{results-details}

\begin{figure}
\vskip -0.2in \hskip -0.4in
\includegraphics[angle=90,width=3.4in,clip]
{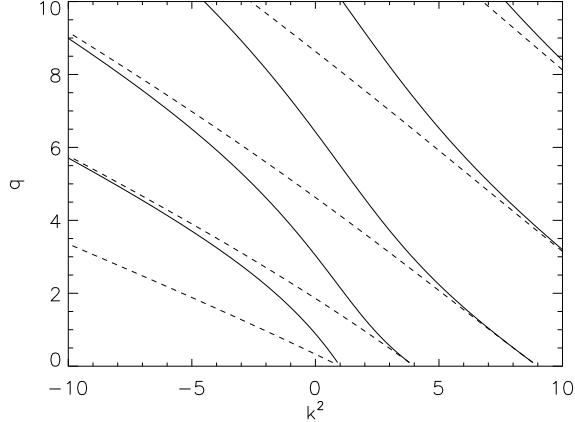}
\vskip -.2in \caption{This figure shows the boundaries of the five
  lowest instability bands for the Mathieu equation using~\eqref{aeq}
  and~\eqref{qeq}.  The lower bound of a given band is the dashed
  curve and the upper bound is the solid curve immediately above it.
  Note that there are no modes of the quantum field with $k^2 < 0$.}
\label{fig-bands}
\end{figure}

As discussed in Section~\ref{3backreaction}, as $\langle \psi^2
\rangle_{\rm ren}$ increases due to parametric amplification the
amplitude of the field damps and its oscillation frequency increases.
Thus, it is useful to define an effective value for $|q|$, which we
call $q_{\rm eff}$, and whose definition is given in Eq.~\eqref{qeff}.
If $q_{\rm eff} \gg 4 k^2 >0 $ for some or all of the modes in an
instability band, the oscillation frequency for the modes is much larger
than that for the inflaton field.  KLS point out that this allows one
to use the Mathieu equation to provide a reasonable description of
what happens since $q_{\rm eff}$ varies slowly compared to the
oscillation time for the modes.  If $q_{\rm eff}$ is small enough then
it seems likely that a description in terms of the Mathieu equation
breaks down from a quantitative, and perhaps also from a qualitative,
point of view.

Our explanations of the details of the damping process are based on
the basic scenario outlined by KLS in which as the damping proceeds,
$q_{\rm eff}$ decreases and there are various instability bands that
cross over a given set of modes.  The damping stops when $q_{\rm eff}$
gets small enough.  The locations of the five lowest instability bands in the $(k^2, q)$ plane are shown in Fig. \ref{fig-bands}.

There are four primary factors that go into our
explanations:
\begin{enumerate}

\item
The lowest instability band, {\it i.e.,} that with the smallest values
of $k$, has the largest value of $\mu_{\rm max}$ and is the widest.
Thus the modes near its center undergo the largest amount of
parametric amplification.

\item
As $q_{\rm eff}$ decreases the tendency is for instability bands to
move to the right along the $k$ axis and for new ones to appear when
$k^2 = \omega_{\rm eff}^2 (a_{\rm eff} - 2q_{\rm eff})$ becomes
positive for part of the band.

\item
Modes near the center of an instability band will continue to undergo
significant parametric amplification (compared with the rest of the
band) until the backreaction on the inflaton field causes $q_{\rm
  eff}$ to decrease enough that these modes are either near the edge
of the instability band (where $\mu$ is small) or no longer in it, due
to the shift of the band along the $k$ axis.

\item
Once the lowest instability band shifts significantly so that the
original modes near the center stop growing rapidly, new modes will
begin to undergo a significant amount of parametric amplification.
However, it takes a while for the contributions of these new modes to
the value of $\langle \psi^2 \rangle_{\rm ren}$ to become significant
compared with that of the modes which were previously near the center
of the first instability band, and have now stopped undergoing
significant parametric amplification.

\end{enumerate}

In what follows we first discuss the case in which no significant
amount of rapid damping occurs.  For the cases in which rapid damping
does occur the pre-rapid damping, rapid damping, and post-rapid
damping phases are discussed separately.

\subsubsection{$g^2 \phi_0^2 \stackrel{<}{_\sim} 1$}

\begin{figure}
\vskip -0.2in \hskip -0.4in
\includegraphics[angle=90,width=3.4in,clip]
{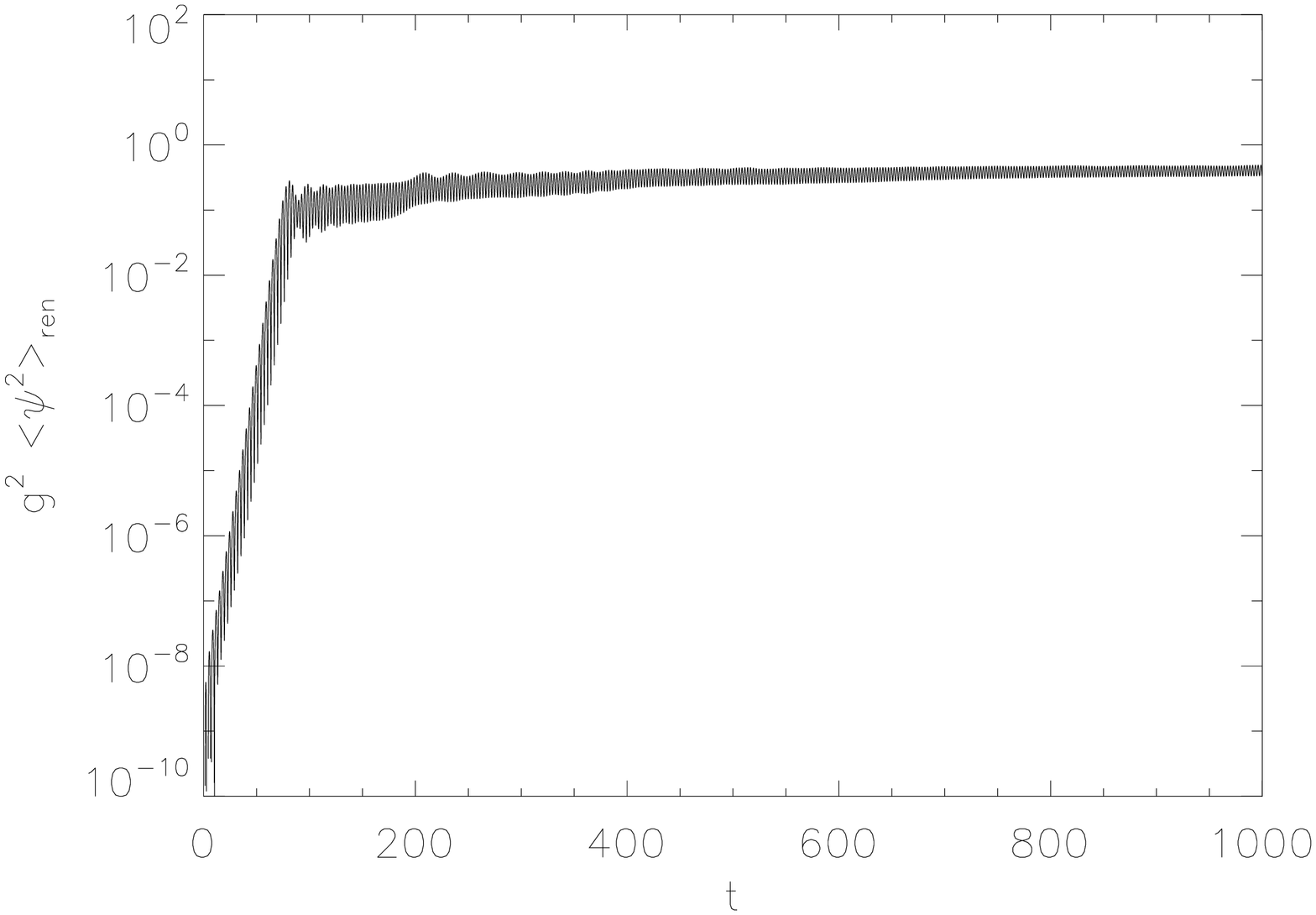}
\includegraphics[angle=90,width=3.4in,clip]
{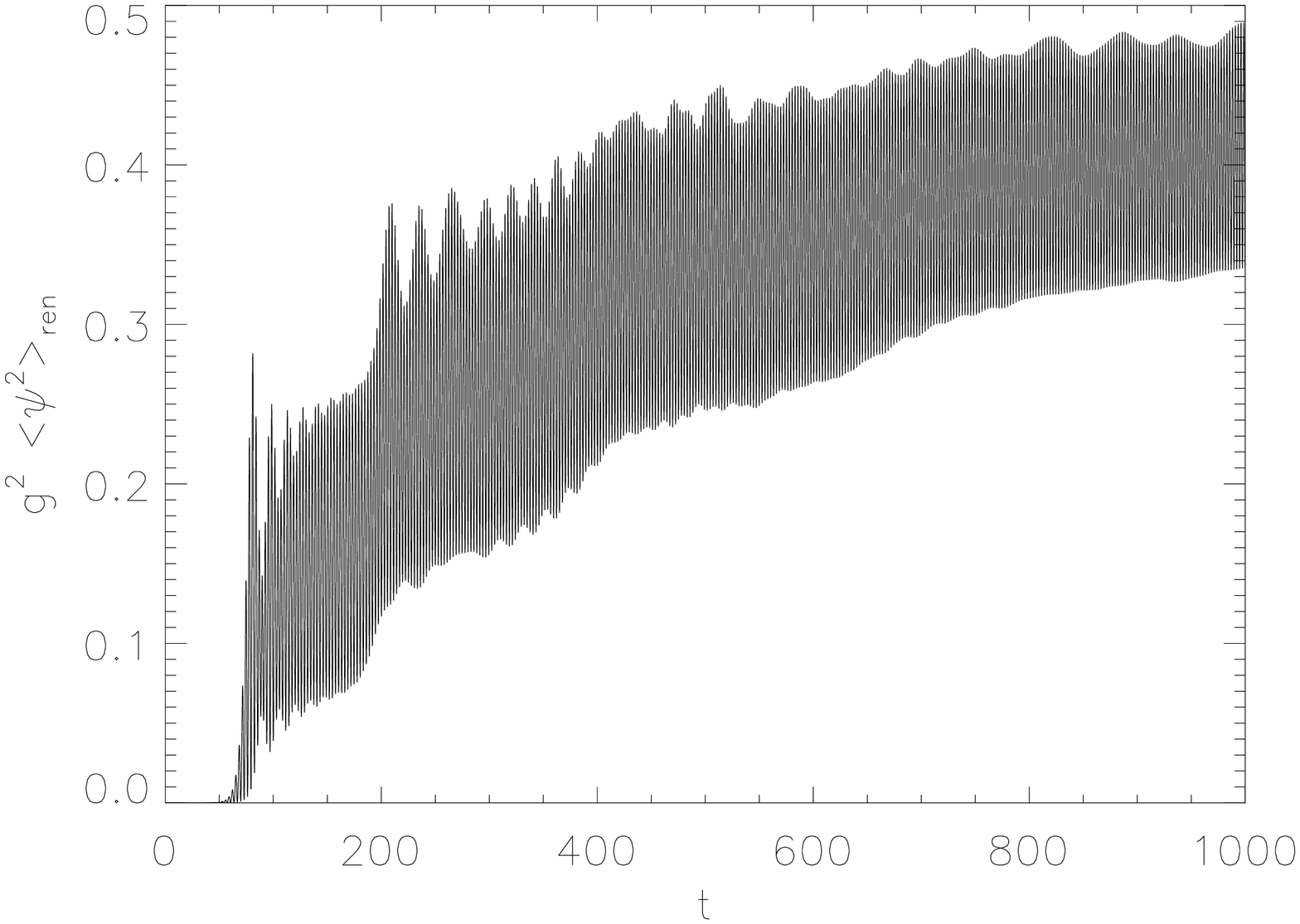}
\vskip -.2in \caption{These plots show $g^2 \langle \psi^2
  \rangle_{\rm ren}$ for $g = 10^{-3}$ and $g^2 \phi_0^2 = 1$.  The
  plots are identical except that the one on the left is plotted on a
  logarithmic scale and the one on the right is plotted on a linear
  scale. }
\label{fig-psi2-gphi2-1}
\end{figure}

\begin{figure}
\vskip -0.2in \hskip -0.4in
\includegraphics[angle=90,width=3.4in,clip]
{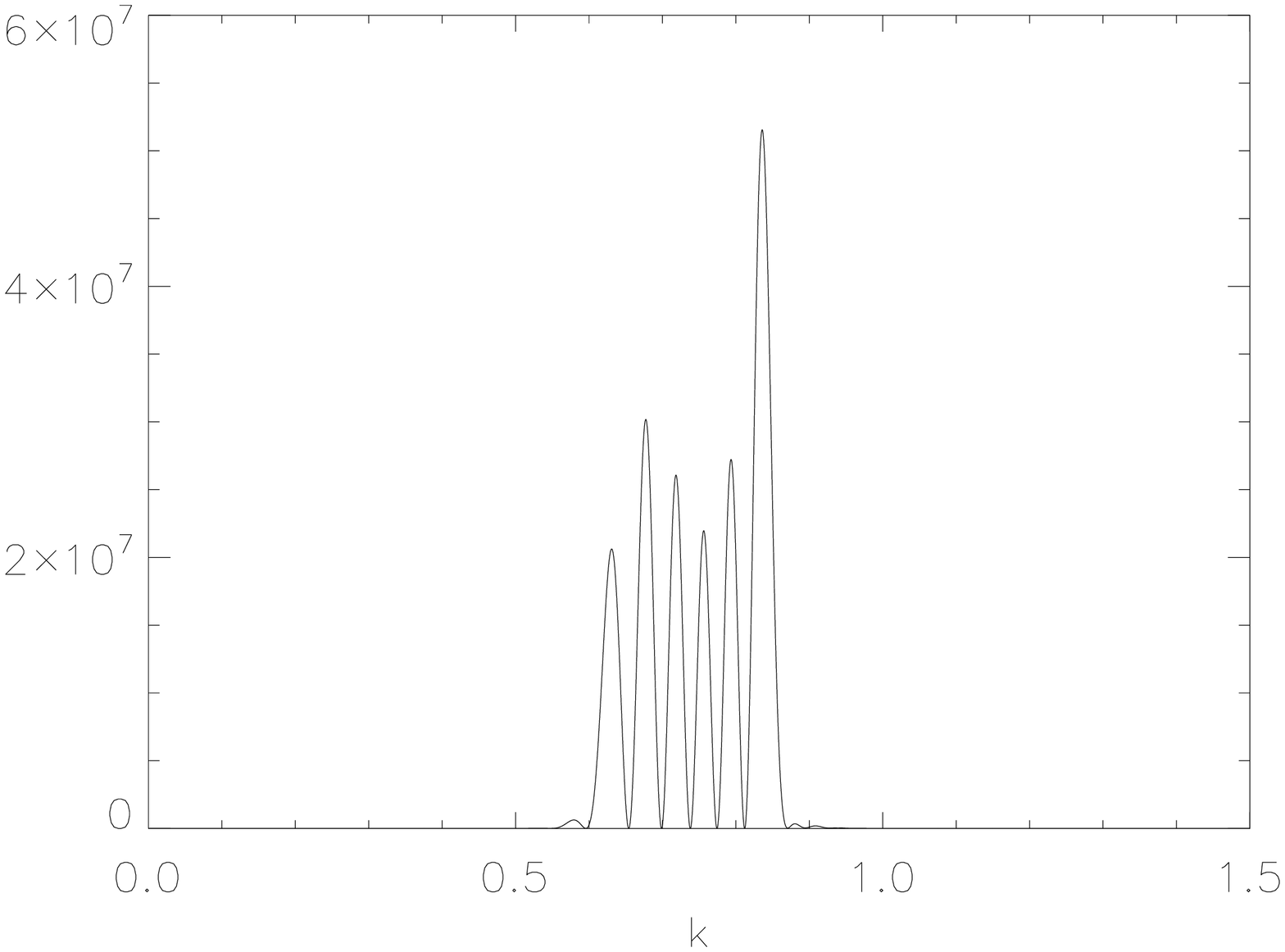}
\includegraphics[angle=90,width=3.4in,clip]
{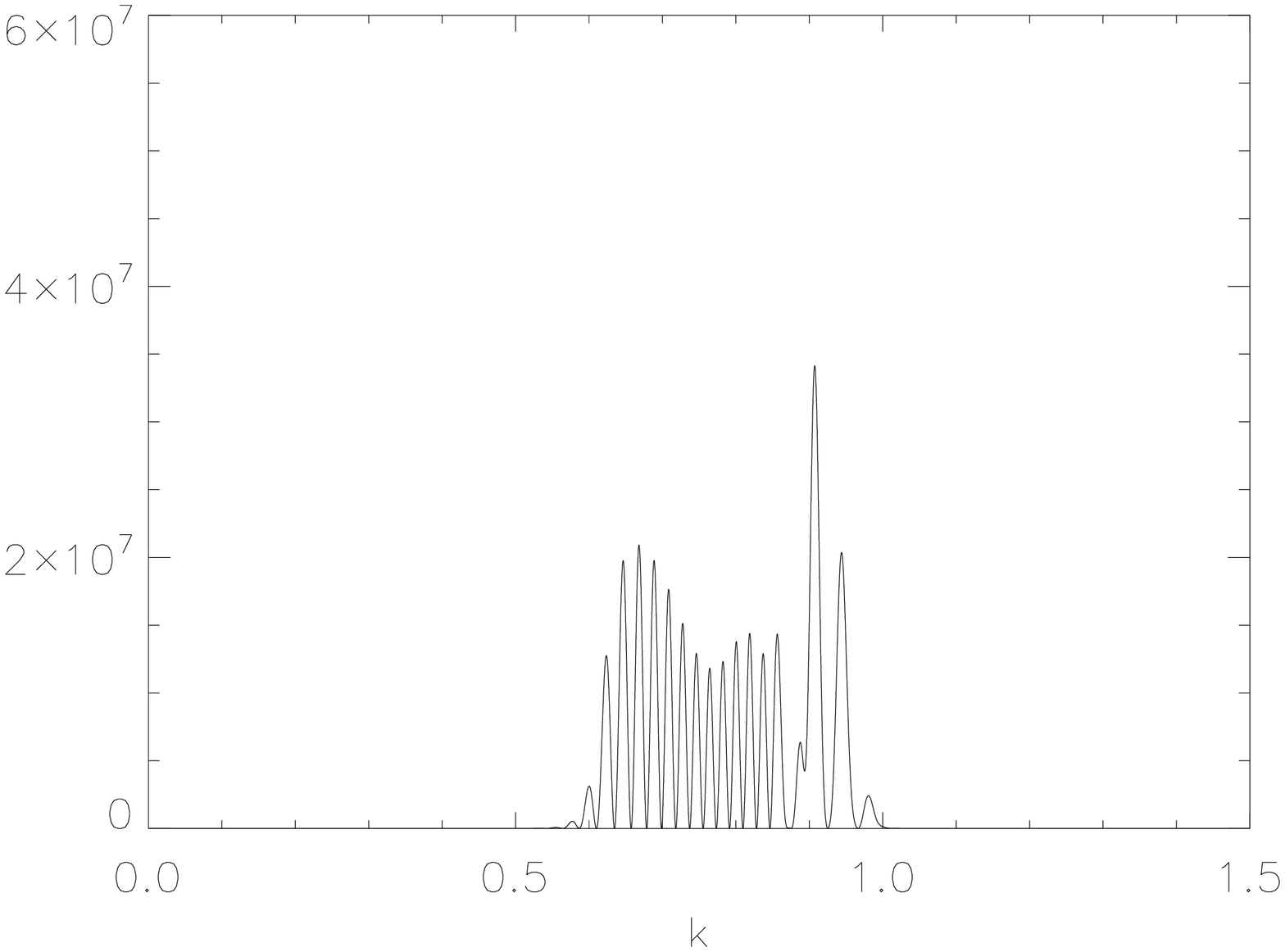}
\vskip -.2in \caption{This figure shows the integrand in
  Eq.~\eqref{psi2ren-a} for the case $g = 10^{-3}$, $g^2 \phi_0^2 =
  1$, for the times $t = 150$ (left) and $t = 250$ (right).  Note that
  at the later time the band of modes that contributes significantly
  to the integral includes modes with larger values of $k$ than the
  band at the earlier time. }
\label{fig-int-gphi2-1}
\end{figure}

KLS predict and we observe that there is no significant amount of
rapid damping for $g^2 \phi_0^2 \stackrel{<}{_\sim} 1$.  However, the
details are different because they consider an expanding universe.  In
the Minkowski spacetime case, as can be seen by comparing the lower
plot of Fig.~1 and the plots in Fig.~\ref{fig-psi2-gphi2-1}, there is actually a
small amount of rapid damping that occurs for $g^2 \phi_0^2 = 1$ once
$g^2 \langle \psi^2 \rangle_{\rm ren} \sim 10^{-2}$.  What appears to
be happening is that modes near the center of the first instability
band cause $\langle \psi^2 \rangle_{\rm ren}$ to grow exponentially
until the band has shifted to the right along the $k$ axis enough so
these modes no longer grow so rapidly.  Because of the exponential
growth of $g^2 \langle \psi^2 \rangle_{\rm ren}$, the damping of the
inflaton field is very rapid once it begins to occur in a significant
way.  However, as $|q| = 0.25$ in this case, the first instability
band is relatively narrow and thus not much damping of the inflaton
field needs to occur for this band to shift by the required amount.
Once it has shifted, new modes begin to undergo significant parametric
amplification and as they start to contribute to $\langle \psi^2
\rangle_{\rm ren}$ the amplitude of the inflaton field continues to
decrease and the position of the instability band continues to move
slowly to the right along the $k$ axis.  One can see evidence of this
shift by comparing plots of the integrand for $\langle \psi^2
\rangle_{\rm ren}$ at two different times as is done in
Fig.~\ref{fig-int-gphi2-1}.

As the inflaton field continues to damp and $q_{\rm eff}$ continues to
decrease, the value of $\mu$ at the center of the band should also be
decreasing.  Thus, the time scale for the new modes undergoing
parametric amplification to contribute significantly to the damping
increases and the rate of damping decreases.


\subsubsection{$g^2 \phi_0^2 \stackrel{>}{_\sim} 2$}

For $g^2 \phi_0^2 \stackrel{>}{_\sim} 2$ rapid damping has occurred in
every case that has been numerically investigated.  However, the time
at which the rapid damping occurs and the details relating to how the
rapid damping occurs and how much damping there is vary widely.  To
understand the reasons for this, it is useful to break the discussion
up into the pre-rapid damping, rapid damping and post-rapid damping
phases.

\subsubsection*{a.  Pre-Rapid Damping Phase}

Since $g^2 \langle \psi^2 \rangle_{\rm ren} \ll 1$ initially for all
cases considered in this paper, there is no significant amount of
backreaction on the inflaton field at first.  Therefore, parametric
amplification of the modes near the center of the first instability
band, {\it i.e.,} the one which encompasses the smallest values of
$k$, occurs and the contribution of these modes
quickly dominates all other contributions to $\langle \psi^2
\rangle_{\rm ren}$. \footnote{Of course, if initially $k^2 >0$ for only a small part of the first instability band,
then the modes in this band may make a comparable or even smaller contribution to $\langle \psi^2 \rangle$
than the modes in the second instability band at early times.}  (As mentioned previously this is in contrast to
the case of an expanding universe where the expansion causes a
decrease in the amplitude of the inflaton field and thus in $q_{\rm
  eff}$.)  As before, $\langle \psi^2 \rangle_{\rm ren}$ grows
exponentially until it becomes of order $10^{-2}$ at which point
significant damping of the inflaton field begins to occur.  What
happens next is observed to depend upon the location on the $k$ axis
of the first instability band.

\begin{enumerate}

\item

If $k^2 < 0$ for a significant fraction of the band and $k^2 > 0$ for
a significant fraction then the inflaton field goes directly into the
rapid damping phase which is discussed next.  A good example of this
is the case $g^2 \phi_0^2 = 10$ which is shown in Fig.~\ref{fig-comp}.
The average value of $g^2 \langle \psi^2 \rangle$ increases
exponentially until it gets large enough to significantly affect the
inflaton field and cause the first phase of rapid damping to occur.
Thus there is no significant amount of gradual damping that occurs
before the rapid damping phase.

\item
If $k^2 > 0$ for most or all of the band then as $q_{\rm eff}$
decreases, the shift in the band causes the modes near the center to
stop increasing rapidly in amplitude.  In that case the exponential
increase in $\langle \psi^2 \rangle_{\rm ren}$ ceases.  At this point
there are two possibilities that have been observed:

\begin{enumerate}

\item
If the decrease in $q_{\rm eff}$ has resulted in $k^2 >0$ for part or
all of an instability band in the $(a,q)$ plane for which previously
$k^2 < 0$, then the modes which are now in that band will begin to
increase in amplitude exponentially and will do so at a faster rate
than modes in what was previously the first instability band.

\item
If the decrease in $q_{\rm eff}$ has not been large enough for a new
instability band to appear, then slow damping will start to occur due
to parametric amplification of those modes that are now near the
center of the first instability band but which were not close to it
before.  As they become large enough to have an effect, $q_{\rm eff}$
will gradually decrease and the slow damping which occurs is of the
type discussed for the $g^2 \phi_0^2 \stackrel{<}{_\sim} 1$ case.
Eventually $q_{\rm eff}$ will become small enough that an instability
band that previously had $k^2 < 0$ will now encompass $k^2 = 0$ and
parametric amplification will occur, again at a faster rate than that
which is occurring in what was previously the first instability band.
The modes will grow exponentially until rapid damping begins.  This is
what happens for $g^2 \phi_0^2 = 35$.  A careful examination of the
first plot in Fig.~1 shows the gradual damping phase followed by the
rapid damping phase.
\end{enumerate}
\end{enumerate}

\subsubsection*{b. Rapid Damping Phase}

In the rapid damping phase, as mentioned previously, KLS point out
that a Minkowski spacetime approximation is valid and they suggest
that the value of $q_{\rm eff}$ should decrease until it reaches about
$1/4$, at which point there are only narrow instability bands, so the
rapid growth of $g^2 \langle \psi^2 \rangle_{\rm ren}$ ceases, as does
the fast damping of the inflaton field.  They predict that the ratio
of the amplitude of the inflaton field after the rapid damping to
that before the rapid damping is
\begin{eqnarray}
\left( \frac{A}{A_0} \right)_{\rm KLS}
 &=& \left(\frac{4}{g^2 A_0^2} \right)^{\frac{1}{4}}
\; ,
\end{eqnarray}
with $A$ the amplitude of the inflaton field $\phi$ after the rapid damping, and $A_0$
the amplitude just before the rapid damping begins.  We consistently
find that after the rapid damping has finished $q_{\rm eff} \stackrel{<}{_\sim}
10^{-2}$.  This is significantly
smaller than the value of $1/4$ that they predict.  As a result, the
amount of damping which the inflaton field undergoes is much larger
than the amount predicted by KLS.  After the rapid damping, the value
of $q_{\rm eff}$ is observed to vary periodically as the envelope and
frequency of the oscillations of the inflaton field change in time.

A careful examination of the plots showing the evolution of $\phi$
indicates that the rapid damping always seems to occur in two
different phases separated by a short time.  After the first phase the
field is damped by a significant amount, but $q_{\rm eff}$ is still
large enough that significant parametric amplification can occur for
modes in the lowest instability band.  As it is occurring, but before
it gets large enough to have a significant effect on $g^2 \langle
\psi^2 \rangle_{\rm ren}$, the inflaton field does not undergo any
more noticeable damping.  Once there is enough of an effect, the
second phase of rapid damping takes place and after it is over,
$q_{\rm eff}$ becomes small enough so that no more significant damping
appears to occur even after a large number of oscillations of the
inflaton field.

\begin{figure}
\vskip -0.2in \hskip -0.4in
\includegraphics[angle=90,width=3.4in,clip]
{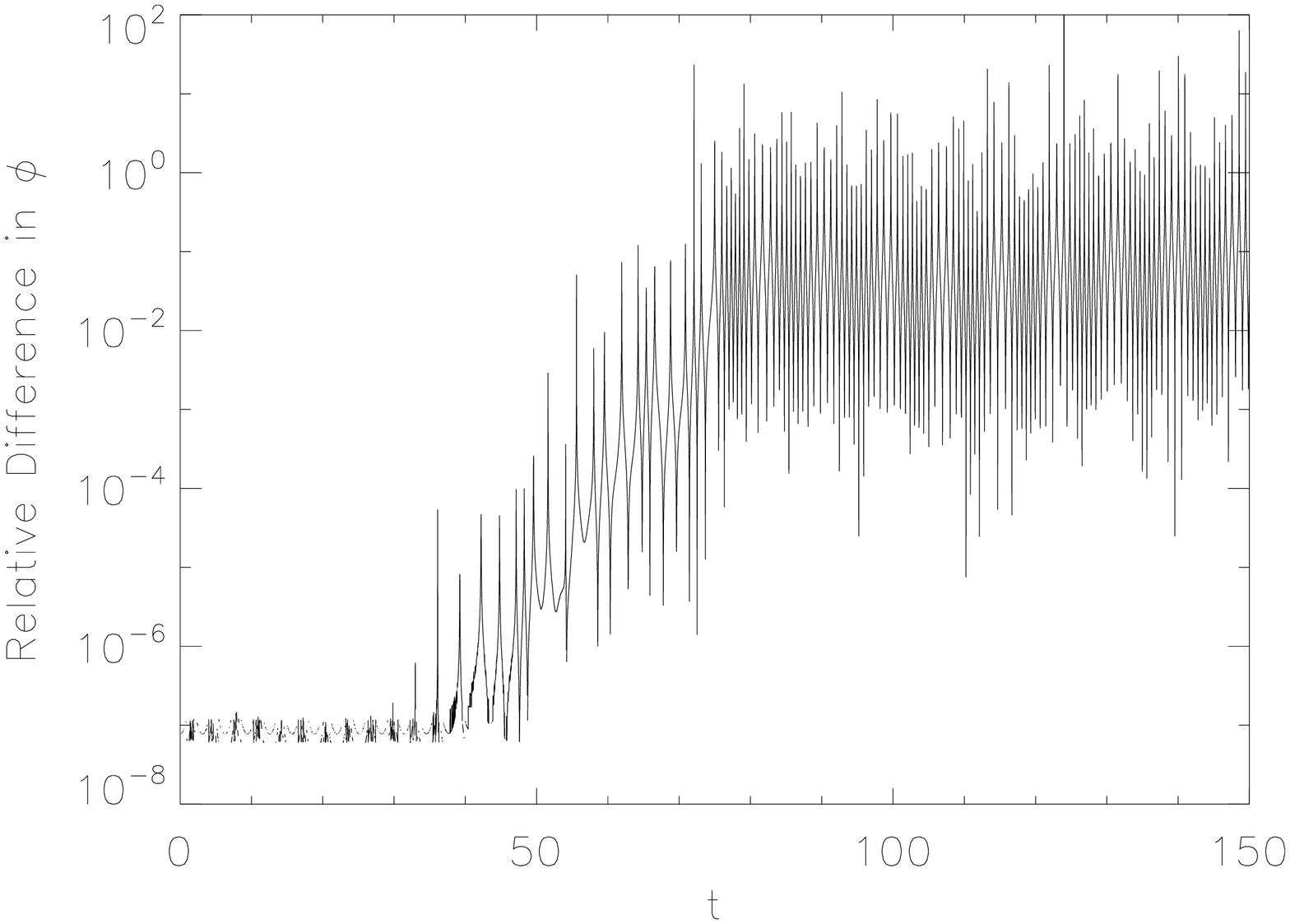}
\includegraphics[angle=90,width=3.4in,clip]
{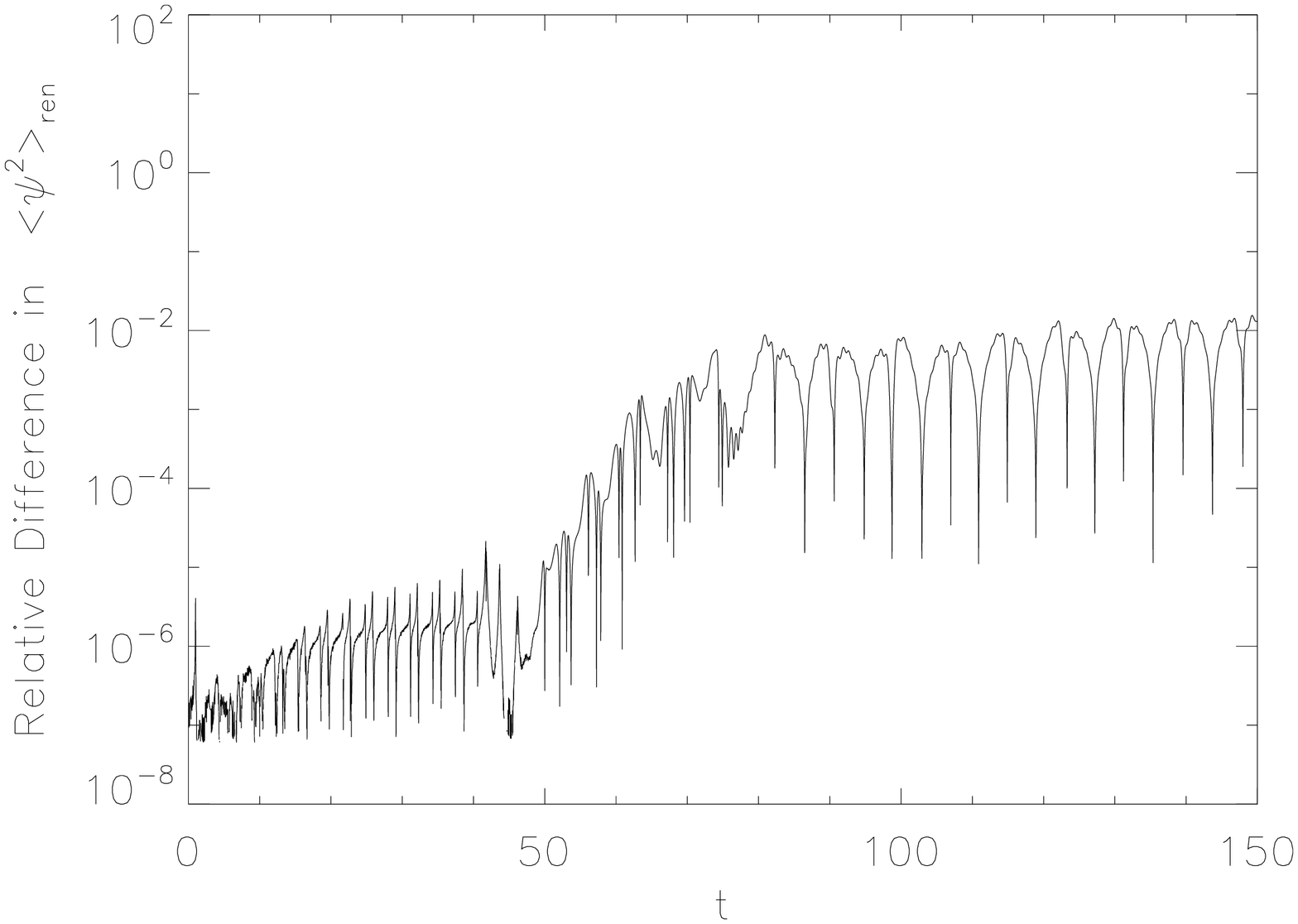}
\vskip -0.2in \hskip -0.4in
\includegraphics[angle=90,width=3.4in,clip]
{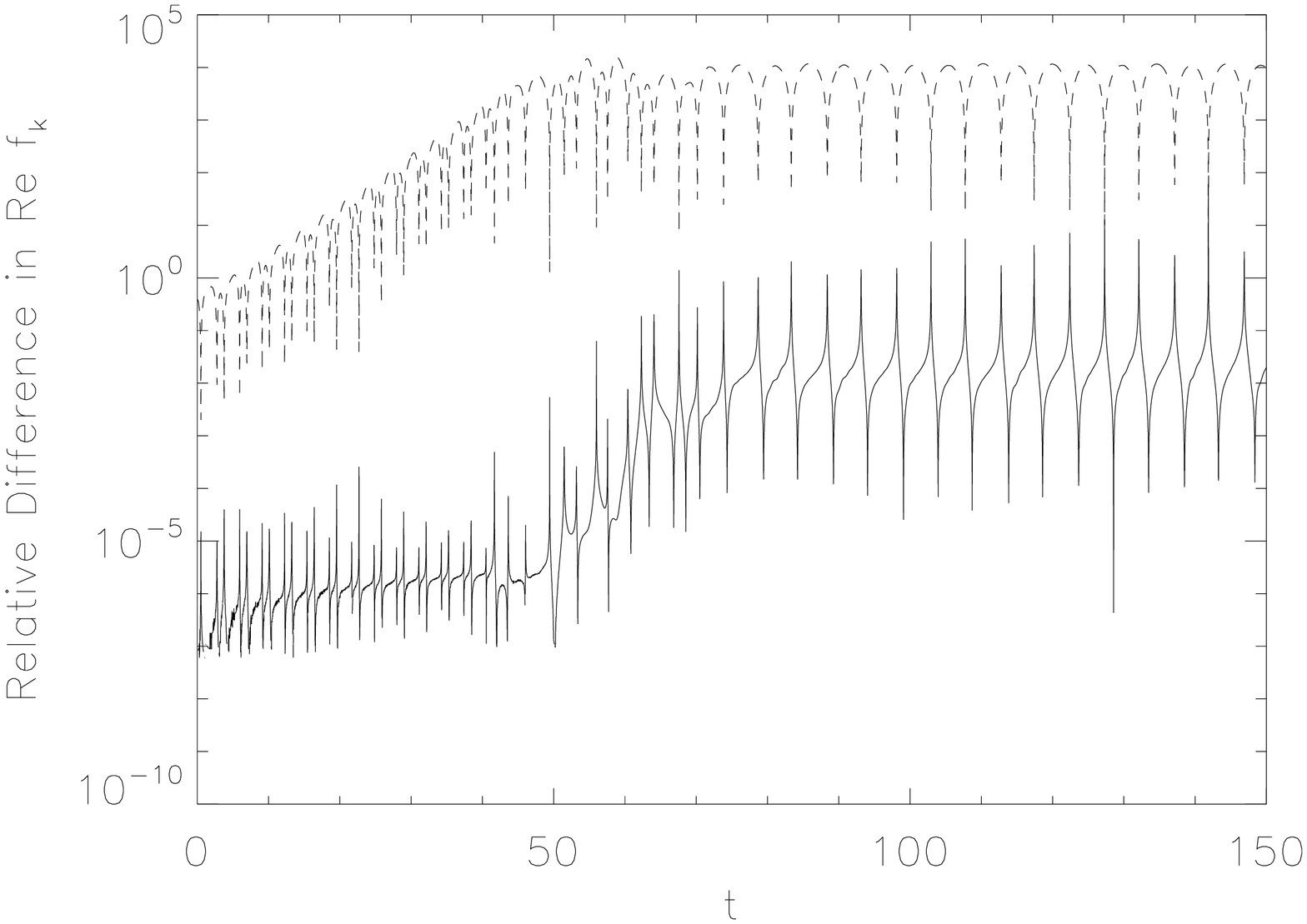}
\vskip -.2in \caption{The time evolution is shown for the case in
  which $g = 10^{-3}$ and $g^2 \phi_0^2 = 10$ for one solution and
  $10(1+10^{-7})$ for the other.  The top plots show the evolution of
  $|\delta \phi/\phi|$ and $|\delta \langle \psi^2 \rangle_{\rm
    ren}/\langle \psi^2 \rangle_{\rm ren}|$.  The bottom plot shows the
  evolution of $|\delta {\rm Re}\,f_k/({\rm Re} \,f_k)|$ for $k =
  0.5$.  The dashed curve corresponds to $|{\rm Re} f_k|$ and is not a relative difference.  Note that
  the growth of the relative difference for each quantity shuts off before its average
  becomes of order unity.  The regions of the plots where the average
  evolves as a straight line with a nonzero slope correspond to times
  when the average relative differences are growing exponentially.  }
\label{fig-diffphipsi1}
\end{figure}

\begin{figure}
\vskip -0.2in \hskip -0.4in
\includegraphics[angle=90,width=3.4in,clip]
{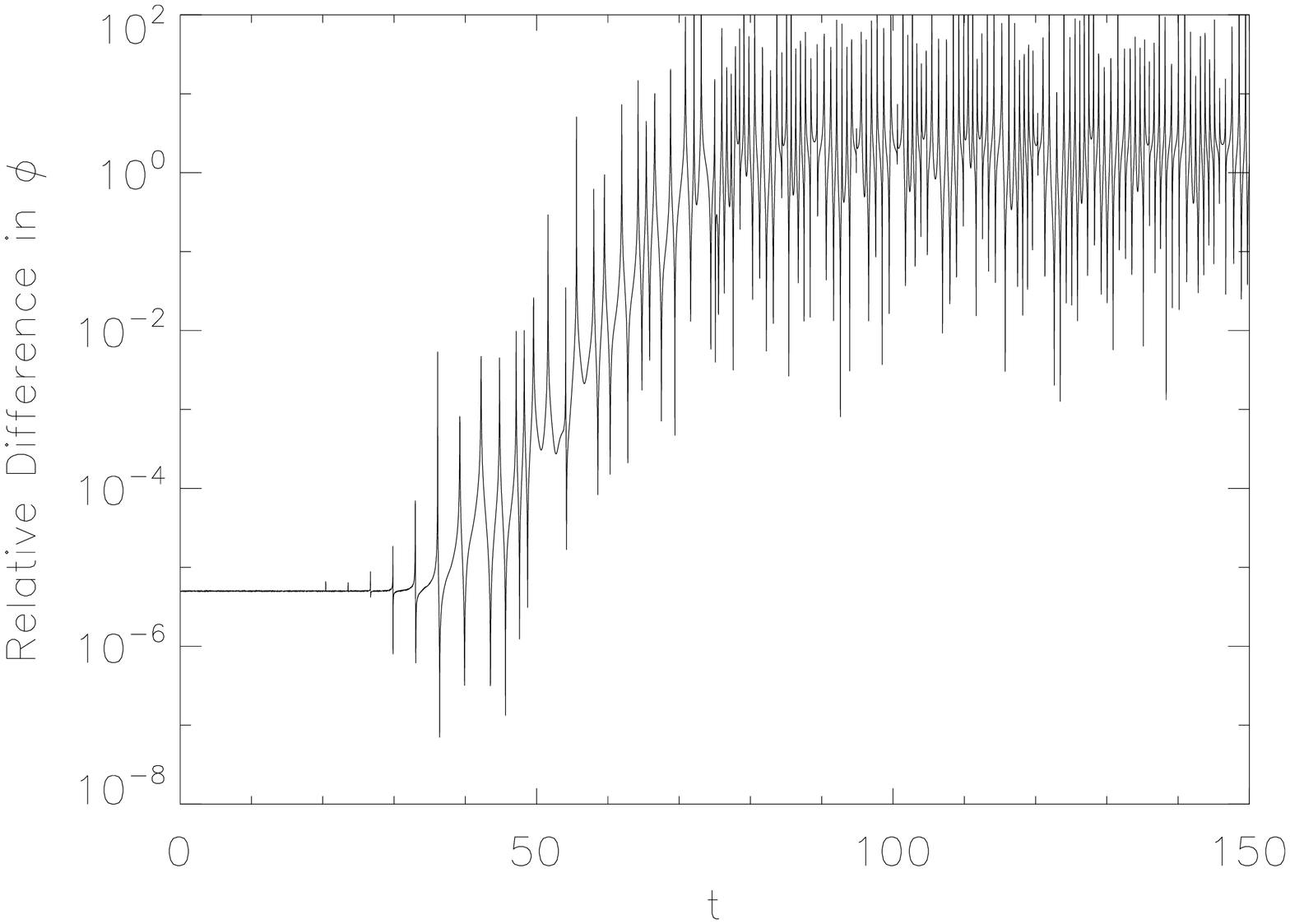}
\includegraphics[angle=90,width=3.4in,clip]
{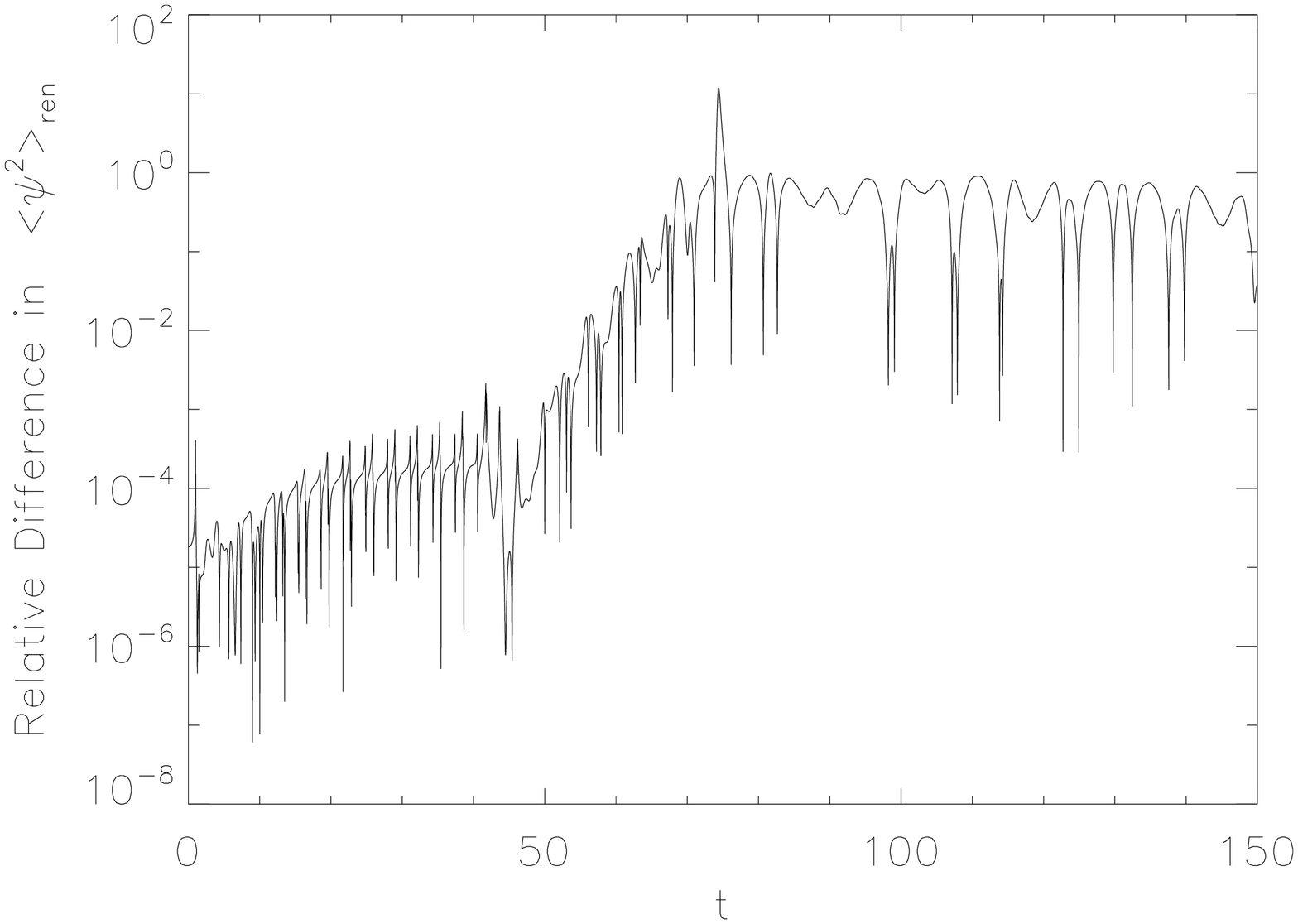}
\vskip -0.2in \hskip -0.4in
\includegraphics[angle=90,width=3.4in,clip]
{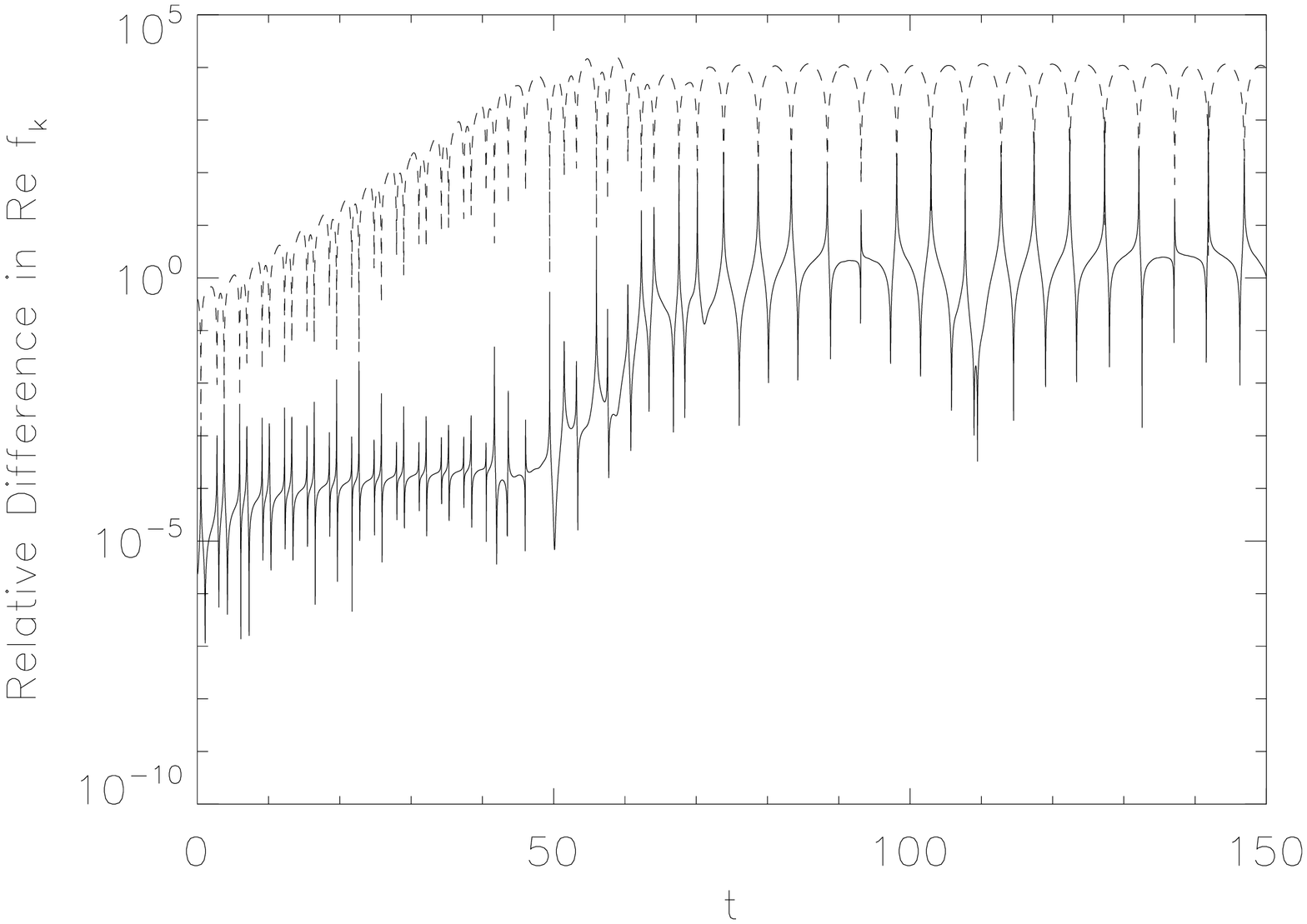}
\vskip -.2in \caption{The time evolution is shown for the case in
  which $g = 10^{-3}$ and $g^2 \phi_0^2 = 10$ for one solution and
  $10(1+10^{-5})$ for the other.  The top plots show the evolution of
  $|\delta \phi/\phi|$, $|\delta \langle \psi^2 \rangle_{\rm
    ren}/\langle \psi^2 \rangle_{\rm ren}|$.  The bottom plot shows the
  evolution of $|\delta {\rm Re}\,f_k/({\rm Re} \,f_k)|$ for $k =
  0.5$.  The dashed curve corresponds to $|{\rm
    Re} f_k|$ and is not a relative difference.  The regions of the plots where the average evolves as
  a straight line with a nonzero slope correspond to times when the
  average relative differences are growing exponentially.  }
\label{fig-diffphipsi2}
\end{figure}

It is during this period, begun by the first rapid damping of the
inflaton field and finished by the second, that the extreme
sensitivity to initial conditions that we have found seems to
manifest.  A useful way to study the sensitivity is to look at the
time evolution of the quantities $\phi$, $g^2 \langle \psi^2 \rangle_{\rm
  ren}$ and $f_k$ for two cases in which the initial data are almost
but not quite identical.  For the cases discussed here, $\phi_0$ is
different by a small amount, which then generates small changes in the
initial values of the modes $f_k$ and thus in $g^2 \langle \psi^2
\rangle_{\rm ren}$.  If the relative differences in the quantities:
$\delta \phi/\phi$, $\delta \langle \psi^2 \rangle_{\rm ren}/\langle
\psi^2 \rangle_{\rm ren}$ and $\delta f_k/f_k$ are plotted as in
Figs.~\ref{fig-diffphipsi1} and \ref{fig-diffphipsi2}, then it is seen
that an exponential increase in their amplitudes takes place between
the time when $g^2 \langle \psi^2 \rangle_{\rm ren} \sim 2$, which is when
the rapid damping begins, and the time when the average of $g^2 \langle \psi^2
\rangle_{\rm ren}$ reaches its maximum value, which is when the rapid
damping ends.

The mechanism for this instability is not completely understood.
However, from Figs.~\ref{fig-diffphipsi1} and \ref{fig-diffphipsi2} it
can be seen that the exponential increase in the average value of
$|\delta \phi/\phi|$ is tied to that of the average value of $|\delta
\langle \psi^2 \rangle/\langle \psi^2 \rangle|$.  Further there is
evidence that the average value of $|\delta f_k/f_k|$ undergoes a
similar exponential increase during the same time period even for
modes which have never undergone parametric amplification.  Finally it
is clear from the plots in Figs.~\ref{fig-comp},
\ref{fig-diffphipsi1}, and \ref{fig-diffphipsi2} that the exponential
increases occur at times when $\phi$, $\langle \psi^2 \rangle_{\rm
  ren}$, and, at least for the mode shown, $f_k$, are not increasing
exponentially.

To understand how this might occur, consider two cases, labeled by the
subscripts $1$ and $2$ such that $\delta \phi = \phi_1 - \phi_2$.
Then the equations for the evolution of $\delta \phi$ and $\delta f_k$
are
\begin{subequations}
\begin{eqnarray}
\delta \ddot \phi &=& - (1 + g^2 \,
\langle \psi_1^2 \rangle_{\rm ren})\, \delta \phi
- g^2 \,\delta \langle \psi^2 \rangle_{\rm ren} \,
\phi_2
\; ,
\label{delphi}
\\
\delta \ddot f_k &=& - (k^2 + g^2 \, \phi_1^2) \,
\delta f_k  - (f_k)_2 \, g^2 \, (\phi_1^2 - \phi_2^2)
\nonumber
\\
& \approx &  - (k^2 + g^2 \, \phi_1^2) \,
\delta f_k + 2 \,(f_k)_2 \, g^2 \, \phi_1 \, \delta \phi
\; .
\label{delfk}
\end{eqnarray}
\end{subequations}
Thus, it is clear that $\delta \langle \psi^2 \rangle$ is a source for
$\delta \phi$ and $\delta \phi$ is a source for $\delta f_k$.  In this
way, one can understand how the exponential increases could be
correlated.

It is not clear what is driving the exponential increase or what stops
it.  However, it is clear that it starts when backreaction effects
from the quantum field on the inflaton field become very important,
which is just before the rapid damping occurs.  It is also clear that
it stops once the rapid damping is finished.  At this point $q_{\rm
  eff} \ll 1$.

This appears to be a different type of sensitivity to initial
conditions than that found by KLS.  Theirs is related to sensitivity
of the evolution of the phase of $\phi$ and the value of the
exponential parameter $\mu$ to absolute changes in the parameter $q$
which are of order unity, but which correspond to small relative
changes in $q$ for large values of $q$.  The sensitivity discussed
here has been observed to occur for $0.5 \stackrel{<}{_\sim} q
\le 25$.  The only larger value of $q$ that was investigated
was $q = 10^4$.  For $q = 10^4$ a sensitivity to initial
conditions was observed which becomes important at earlier times than
the ones we found for smaller values of $q$.  Thus it seems likely
that this is the sensitivity discussed by KLS.  The sensitivity to initial conditions that
we have found results in significant changes in the evolution of the
inflaton field for changes in $q$ which in some cases have been observed to be
less than one part in $10^8$, for values of $q$ of order unity.

\subsubsection*{c.  Post-Rapid Damping Phase}

As mentioned in the Introduction, the model considered here does not
include quantum fluctuations of the inflaton field and also does not
include interactions between the created particles or between those
particles and quantum fluctuations of the inflaton field.  It would
naturally be expected that scattering due to these interactions would
become important at some point and that they would lead to the
thermalization of the particles.  Various studies which have taken
such interactions into
account~\cite{tkachev,prokopec-roos,berges-serreau,pfkp,dfkpp} indicate
that scattering effects will become important at intermediate times.
In some cases this may occur before the rapid damping phase has
ended~\cite{prokopec-roos,dfkpp}.  Once the scattering becomes important the
evolution should change significantly from that found using our
model.

Nevertheless it is of some interest to see what happens in our model
at late times because of the insight it gives into the effects of
particle production in the context of the semiclassical approximation.
First, since quantum fluctuations of the inflaton field are ignored,
one would expect that the damping of the inflaton field is due to the
production of particles of the quantum fields.  However, in the
semiclassical approximation when the classical field is rapidly
varying, different definitions of particle number can give different
answers~\cite{b-d}.  Even if one has a good definition, the particles
may not behave like ordinary particles.  This is the case here as can
be seen in the right hand plot of Fig.~\ref{fig-energy}.  Recall that
in Minkowski spacetime the total energy density is conserved.  Then it
is clear from the plot that after the rapid damping phase, most of the
energy is transferred back and forth between the inflaton field and
the quantum field.  Thus, the created particles have an energy density
that is very different from what they would have if the inflaton field
was slowly varying.

\begin{figure}
\vskip -0.2in \hskip -0.4in
\includegraphics[angle=90,width=3.4in,clip]
{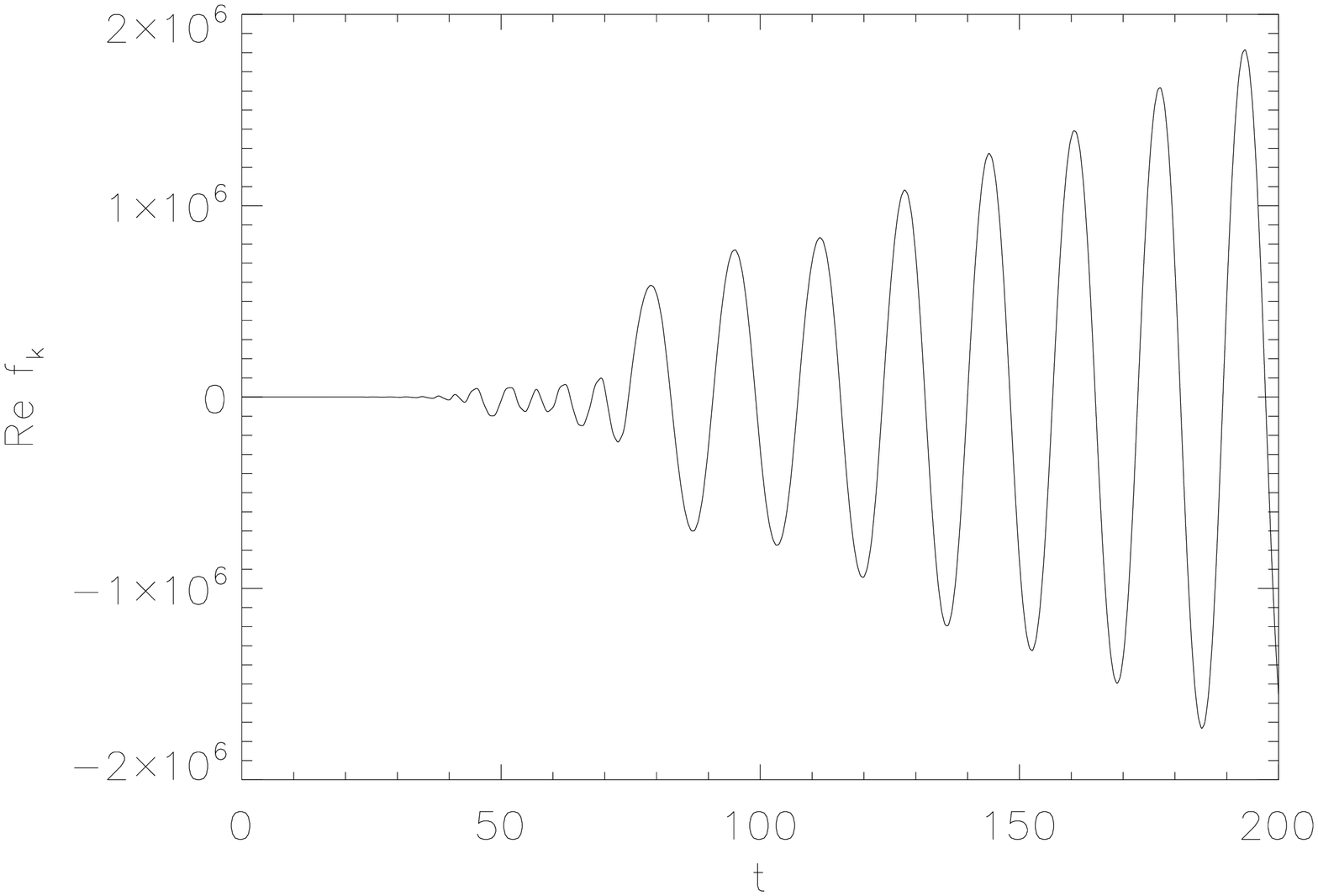}
\includegraphics[angle=90,width=3.4in,clip]
{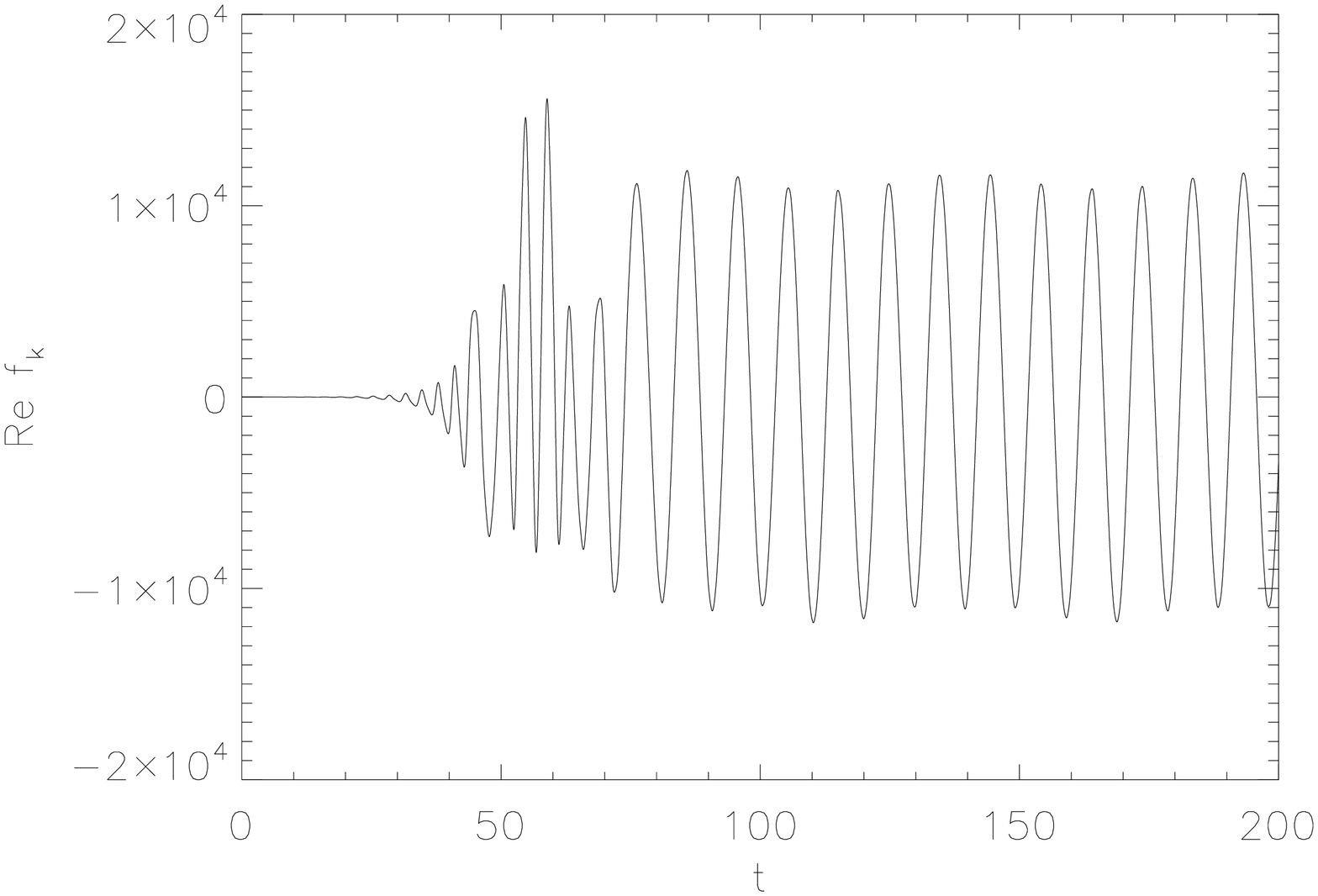}
\vskip -.2in \caption{Shown is the real part of two mode functions
  $f_k$ for the case $g = 10^{-3}$, $g^2 \phi_0^2 = 10$.  For the
  left plot $k = 10^{-4}$ and for the right plot $k = 0.5$. }
\label{fig-modes}
\end{figure}

Also as mentioned previously, damping of the inflaton field is
observed to be negligible for a large number of oscillations after the
rapid damping phase has finished.  However, one would expect that
since the quantum fields are massless, particle production should
continue to occur~\cite{ford}.  Examination of the behaviors of modes
with various frequencies shows that there are some modes which do
continue to grow during the post-rapid damping phase.  One would
expect that these modes would have $k^2 \sim g^2 A^2/\omega_{\rm
  eff}^2 = 4 q_{\rm eff}$.  The left hand plot of Fig.~\ref{fig-modes}
shows that such growth does occur.  However, it does not appear to be
exponential and this may be the reason that these modes have not grown
enough to contribute significantly to $g^2 \langle \psi^2 \rangle_{\rm
  ren}$, even after a large number of oscillations of the inflaton
field.  For other modes, such as that in the plot on the right of
Fig.~\ref{fig-modes}, significant growth ceases after the rapid
damping of the inflaton field.  Finally, as would be expected for
modes with large values of $k$, no significant growth in the amplitude
of the oscillations was observed to occur.

\section{Discussion and conclusions}
\label{discussion}

We have investigated preheating in chaotic inflation using the large
$N$ expansion for a classical inflaton field coupled to $N$ identical
massless quantum scalar fields in a Minkowski spacetime background.
The inflaton field is minimally coupled to the gravitational field.

When backreaction effects on the inflaton field are ignored the mode
equations for the quantum fields are Mathieu equations.  We have shown
that, even though parametric amplification occurs in an infinite
number of bands of modes with arbitrarily large frequencies, at any
finite time quantities such as $\langle \psi^2 \rangle_{\rm ren}$ and
$\langle T_{\mu \nu} \rangle_{\rm ren}$ are finite due to the fact
that the widths of the bands become negligible at very high
frequencies and the maximum value of the exponential parameter $\mu$
in a band becomes very small at very high frequencies.

We have investigated the backreaction effects of the quantum fields on
the inflaton field in detail by numerically solving the full set of
equations that couple the inflaton field to the quantum field.  These
are the first such solutions to these equations for the model we are
considering.  We have found that, in agreement with the prediction by
KLS, there is a period of rapid damping of the inflaton field if $g^2
\phi_0^2 \stackrel{>}{_\sim} 2$ and none otherwise.  In both cases the
details of the damping can be understood through an analysis similar
to that done by KLS; this analysis involves studying the way in which
the instability bands change as the inflaton field damps.  For $g^2
\phi_0^2 \stackrel{>}{_\sim} 2$ the situation depends upon the
locations of the instability bands, and therefore, there are different
ways in which the damping proceeds.  For $g^2 \phi_0^2
\stackrel{<}{_\sim} 1$ there is a brief period in which the amplitude
of the inflaton field is rapidly damped but only by a very small
amount.  This is followed by a long period of slow damping.  Thus, we
distinguish this from the cases in which the inflaton field is rapidly
damped by a significant amount.

The analysis by KLS shows that significant differences in the
evolution occur for realistic values of the initial conditions and
parameters when the expansion of the universe is taken into account.
However, in cases where rapid damping of the inflaton field occurs, it
is possible to neglect this expansion during the rapid damping phase.
We find that during the rapid damping phase there is more damping of
the inflaton field than was predicted originally by KLS.  We also find
that the rapid damping occurs in two bursts separated by a period of
time which is generally longer than the time it takes for each period
of rapid damping to occur.  Finally, we find that there is an extreme
sensitivity to initial conditions which manifests in significant
changes in the evolution of the inflaton field and the modes of the
quantum fields during the period of rapid damping.  This can occur for
changes in the initial amplitude of the inflaton field as small as, or
even smaller than, one part in $10^8$.  Since this sensitivity occurs for values of $g^2 \phi_0^2
\stackrel{>}{_\sim}2$, it would seem to be different than the one
discussed by KLS, which only occurs for large values of $g^2
\phi_0^2$.

\vskip .1cm
\centerline{{\it Acknowledgments}}
\vskip .1cm

P.R.A. would like to thank L. Ford and B. L. Hu for helpful
conversations.  We would like to thank L. Kofman, I. Lawrie, A. Linde, and A.A. Starobinsky
for helpful comments regarding the manuscript.  P.R.A. and
C.M.-P. would like to thank the T8 group at Los Alamos National
Laboratory for hospitality.  They would also like to thank the
organizers of the Peyresq 5 meeting for hospitality.
 P.R.A.  would like to thank the
Racah Institute of Physics at Hebrew University, the Gravitation Group
at the University of Maryland, and the Department of Theoretical
Physics at the Universidad de Valencia for hospitality. P.R.A.
acknowledges the Einstein Center at Hebrew University, the Forchheimer
Foundation, and the Spanish Ministerio de Educaci\'on y Ciencia for
financial support.  C.M.-P. would like to thank the Department of
Theoretical Physics at the Universidad de Valencia for hospitality and
the Leverhulme Trust for support under a Leverhulme Research
Fellowship RFG/10709.  This research has been partially supported by
Grant No. PHY-0070981, No. PHY-0555617, and No. PHY-0556292 from the National Science
Foundation.  Some of the numerical computations were performed on the
Wake Forest University DEAC Cluster with support from an IBM SUR grant
and the Wake Forest University IS Department.  Computational results
were supported by storage hardware awarded to Wake Forest University
through an IBM SUR grant.

\appendix

\section{Renormalization and conservation of the energy-momentum tensor}
\label{app-tmunu}

In this appendix the renormalization and conservation of the
energy-momentum tensor for the system we are considering are
discussed.  As discussed in Section II, at leading order in a large
$N$ expansion, the system is equivalent to a massive classical scalar
field coupled to a single massless quantum field.  That in turn is
equivalent to the system discussed in Ref.~\cite{amr}, if the scale
factor, $a(t)$, is set equal to one, the coupling of the inflaton
field to the scalar curvature is set to zero, the coupling constant
$\lambda$ is set equal to $2 g^2$, the mass of the quantum field is
set to zero, and the $\lambda \phi^3/3!$ term in the equation of
motion for the $\phi$ field is dropped.  Because of this, we will use
several results from Ref.~\cite{amr} in what follows.

We start with the quantum field theory in terms of bare parameters.
The bare equations of motion for the homogeneous classical inflaton
field and the quantum field modes are given in
Eqs.~\eqref{motion-phi-2} and~\eqref{mode-eq-1} and an expression for
the quantity $\langle \psi^2\rangle_{\rm B}$ is given in
Eq.~\eqref{bare-psi-squared}.  It is assumed throughout that (i) the
inflaton field is homogeneous and thus, depends only on the time $t$
and that (ii) the quantum field $\psi$ is in a homogeneous and
isotropic state so that $\langle \psi^2 \rangle$ depends only on time
and $\langle T_{\mu \nu} \rangle$ when expressed in terms of the usual
Cartesian components, is diagonal and depends only on time.

The total bare energy-momentum tensor of the system is given
by~\cite{amr}
\begin{subequations}
\begin{eqnarray}
T_{\mu \nu}^{B} &=& T_{\mu \nu}^{C,B} + \langle T_{\mu \nu} \rangle^{Q,B}
\; ,
\label{tmunu-total-bare}
\\
T_{\mu \nu}^{C,B}
&=&
\partial_\mu  \phi  \partial_\nu  \phi
- \frac{1}{2} \eta_{\mu \nu}\eta^{\sigma \tau}
\partial_\sigma  \phi  \partial_\tau  \phi
-\frac{m^2_B}{2}
\eta_{\mu \nu}  \phi^2
\; ,
\label{tmunu-classical-bare}
\\
\langle T_{\mu \nu} \rangle^{Q,B}
&=&
(1- 2 \xi_B)
\langle \partial_\mu   \psi  \partial_\nu  \psi  \rangle
+\left( 2 \xi_B - \frac{1}{2}\right)
\eta_{\mu \nu}\eta^{\sigma \tau}
\langle \partial_\sigma  \psi  \partial_\tau   \psi  \rangle
-2 \xi_B
\langle   \psi  \partial_\mu\partial_\nu  \psi  \rangle
\nonumber
\\
&+&
 2 \xi_B \eta_{\mu \nu}
 \langle  \psi \Box   \psi  \rangle
-\frac{g^2_B \phi^2}{2}
\eta_{\mu \nu} \langle \psi^2 \rangle
\; .
\label{quantum-tmunu-bare}
\end{eqnarray}
\label{total-tmunu-bare}
\end{subequations}
We first show that the total energy-momentum tensor is covariantly
conserved.  Since it is diagonal and depends only on time, the
conservation equation in Minkowski spacetime is trivially satisfied
for all of the components of $T_{\mu \nu}^{B}$, except for $T^B_{tt}$.
This component has the explicit form
\begin{subequations}
\begin{eqnarray}
T_{tt}^{C,B}
&=&
\frac{1}{2}\dot \phi^2+ \frac{m^2_B}{2}\phi^2
\; ,
\label{t00-class-bare}
\\
\langle T_{tt}\rangle^{Q,B}
&=&
\frac{1}{4 \pi^2} \int_{0}^{+\infty}   {\mathrm{d}}k \; k^2
\left[
|\dot f_k|^2+ (k^2  + g^2_B \phi^2) |f_k|^2
\right]
\; .
\label{t00-q-bare-bare}
\end{eqnarray}
\end{subequations}
The conservation condition is
\begin{eqnarray}
\partial_t \; T_{tt}^B  &=& 0
\; .
\label{conservation-bare}
\end{eqnarray}
It is easy to show, making use of the equations of motion for
$\phi(t)$ and $f_k(t)$, that
\begin{subequations}
\begin{eqnarray}
\partial_t  \; T_{tt}^{C,B}
&=& - {g^2_B} \phi \; \dot\phi \; {\langle \psi^2 \rangle}_{B}
\; ,
\label{partials-c-bare}
\\
\partial_t \; \langle T_{tt}\rangle^{Q,B}
&=&  {g^2_B} \phi \; \dot\phi \; {\langle \psi^2 \rangle}_{B}
\; ,
\label{partials-b-bare}
\end{eqnarray}
\end{subequations}
with ${\langle \psi^2 \rangle}_{B}$ given by
Eq.~\eqref{bare-psi-squared}. This implies that the total bare
energy-momentum tensor is covariantly conserved.

Renormalization is accomplished through the method of adiabatic
regularization~\cite{parker,zel-star,par-ful,bunch}.  The details are
given in Ref.~\cite{amr}.  The result for $\langle \psi^2 \rangle_B$ is
\begin{subequations}
\begin{eqnarray}
{\langle \psi^2 \rangle}_{\rm ren} &=& \langle \psi^2 \rangle_{\rm B}
- \langle \psi^2 \rangle_{\rm ad}
\label{bare-psi-squared-ren}
\\
&=& \frac{1}{2 \pi^2} \int_{0}^{\epsilon} {\rm d} k \; k^2
\;\left( \left| f_k(t) \right|^2 - \frac{1}{2k} \right)
+
\frac{1}{2 \pi^2} \int_{\epsilon}^{+\infty} {\rm d} k \; k^2
\;\left( \left| f_k(t) \right|^2 - \frac{1}{2k}
+ \frac{g^2_R \phi^2}{4k^3} \right) + \langle \psi^2 \rangle_{\rm an}
\; ,
\nonumber
\\
 \langle \psi^2 \rangle_{\rm an} &=& - \frac{g^2 \phi^2}{8 \pi^2}
\left[1 - \log \left( \frac{2 \epsilon }{M}\right)
\right]
\; .
\end{eqnarray}
\label{psireneq}
\end{subequations}
For the energy-momentum tensor
\begin{subequations}
\begin{eqnarray}
{\langle T_{tt} \rangle}_{\rm ren} &=&
T_{tt}^C + \langle T_{tt} \rangle^Q_{\rm ren}
= T_{tt}^C +  \langle T_{tt} \rangle^Q_B - \langle T_{tt} \rangle^Q_{\rm ad}
\; ,
\label{tmunu-ren-appendix}
\end{eqnarray}
with
\begin{eqnarray}
T_{tt}^C
&=&
\frac{1}{2}\dot \phi^2+ \frac{m^2}{2}\phi^2
\; ,
\label{t00-class}
\\
\langle T_{tt}\rangle^Q_{\rm ren}
&=&
\frac{1}{4 \pi^2} \int_{0}^{+\infty}   {\mathrm{d}}k \; k^2
\left[
|\dot f_k|^2+ (k^2  + g^2 \phi^2) |f_k|^2
\right]
\nonumber \\
  &  &  -
\frac{1}{4 \pi^2} \int_{0}^{\epsilon}   {\mathrm{d}}k \; k^2
\left( k + \frac{g^2 \phi^2}{2k} \right)  \nonumber \\
  &  &  -
\frac{1}{4 \pi^2} \int_{\epsilon}^{+\infty}   {\mathrm{d}}k \; k^2
\left( k + \frac{g^2 \phi^2}{2k}
-
\frac{g^4 \phi^4}{8k^3}
\right)
 + \langle T_{tt} \rangle_{\rm an}^Q
\; ,
 \label{t00-quant}
\\
 \langle T_{tt}\rangle_{\rm an}^{Q} &=&
-\frac{g^4 \phi^4}{32 \pi^2}
\left[1 -
\log \left( \frac{2 \varepsilon }{M}\right)
\right]
\; .
\end{eqnarray}
\label{tabreneq}
\end{subequations}
We note that in Minkowski spacetime the bare and the adiabatic energy
density of the quantum field do not depend on the value of the
coupling $\xi$ of the quantum field to the scalar curvature. Thus, the
renormalized value of the energy density does not depend on $\xi$.

Using Eqs.~\eqref{tabreneq},~\eqref{psireneq}, and the equations of
motion for $\phi$ and $f_k$, it is easy to derive the following
identities:
\begin{subequations}
\begin{eqnarray}
\partial_t \; T_{tt}^C
&=& - {g^2} \phi \; \dot\phi \; {\langle \psi^2 \rangle}_{\rm ren}
\; ,
\label{partials-c}
\\
\partial_t \; \langle T_{tt}\rangle^Q_{\rm B}
&=&  {g^2} \phi \; \dot\phi \; {\langle \psi^2 \rangle}_{\rm B}
\; ,
\label{partials-b}
\\
\partial_t \; \langle T_{tt}\rangle^Q_{\rm ad}
&=&  {g^2} \phi \; \dot\phi \; {\langle \psi^2 \rangle}_{\rm ad}  \;.
\label{partials-ad}
\end{eqnarray}
\end{subequations}
They can be used to show that
\begin{eqnarray}
\partial_t \; \langle T_{tt} \rangle_{\rm ren} &=&
\partial_t \; \langle T_{tt} \rangle^C
+
\partial_t \; \langle T_{tt} \rangle^Q_{\rm B} -
\partial_t \; \langle T_{tt} \rangle^Q_{\rm ad}
\nonumber
\\
&=&
{g^2 \phi \dot\phi \;
\left[ - {\langle \psi^2 \rangle}_{\rm ren}
+ {\langle \psi^2 \rangle}_{\rm B}
- {\langle \psi^2 \rangle}_{\rm ad}\right] }
=0
\; .
\label{con-tmunu-total-ren}
\end{eqnarray}

\end{document}